\documentclass[a4paper,11pt]{article}

\usepackage[dvips]{graphicx}
\usepackage{amsfonts, color, float,  amsmath}
\usepackage{amssymb}\usepackage{subfigure}
\usepackage{url}
\usepackage{array}

\usepackage{slashed}
\usepackage[dvipsnames]{xcolor}  

\usepackage[small,bf,sf]{caption}  
\usepackage{microtype}
\usepackage{cite}
\usepackage{authblk}

\usepackage[bottom]{footmisc}  
\usepackage{booktabs}   
\usepackage{hyperref}

\usepackage{colortbl}
\definecolor{myblue}{rgb}{0.8,0.85,1}  
\definecolor{light-gray}{gray}{0.95}

\hypersetup{colorlinks=true,
	citecolor=ForestGreen,
	anchorcolor=CornflowerBlue,
	linkcolor=NavyBlue,
	urlcolor=Mahogany,
	linkbordercolor={1 1 0}
}

\allowdisplaybreaks

\newcommand{\ah}{\hat a}
\newcommand{\MSb}{{\overline{\rm MS}}}

\newcommand{\tautohad}{\tau\to {\rm (hadrons)}+\nu_\tau}
\newcommand{\epem}{e^+e^-\to {\rm (hadrons)}}
\newcommand{\lambdabar}{{\mkern0.75mu\mathchar '26\mkern -9.75mu\lambda}}

\def\beq {\begin{equation}}
\def\eeq {\end{equation}}
\def\bea {\begin{eqnarray}}
\def\eea {\end{eqnarray}}

\def\nn {\nonumber}

\def\epem{e^+e^-\to(\mbox{hadrons})}

\AtBeginDocument{}

\title{\LARGE {\bf \sffamily \boldmath Renormalons in integrated spectral function moments and \texorpdfstring{$\alpha_s$}{alphas} extractions}}

\author{D. Boito}
\author{F. Oliani\vspace{0.3cm}}

\affil{\it  Instituto de F\'isica de S\~ao Carlos, Universidade de S\~ao Paulo, CP 369, 13560-970, S\~ao
Carlos, SP, Brazil\vspace{0.3cm}}

\date{}

\begin{document}
\begin{flushright}
  {\small \today}
\end{flushright}

\vspace*{-0.7cm}
\begingroup
\let\newpage\relax
\maketitle
\endgroup
\date{}

\vspace*{-1.0cm}
\begin{abstract}
  \noindent
  Precise extractions of $\alpha_s$ from $\tautohad$ and  from $\epem$ below the charm threshold rely
on finite energy sum rules (FESR) where the experimental side is given by integrated spectral function moments. 
Here we study the renormalons that appear in the Borel 
transform of polynomial moments in the large-$\beta_0$ limit and in full QCD. In large-$\beta_0$, we establish a direct connection between the renormalons and the perturbative behaviour of  moments often employed in the literature. 
The  leading IR singularity is particularly 
prominent and is behind the fate of moments whose perturbative series are unstable,
 while those with good perturbative behaviour benefit from  partial cancellations of renormalon singularities.
The conclusions can be extended to  QCD through a convenient scheme transformation to the $C$-scheme together with the use of a modified Borel transform which make the results  particularly simple;  the leading IR singularity becomes a simple pole, as in  large-$\beta_0$.  Finally, 
for the moments that display good perturbative behaviour, we discuss an  optimized truncation based on renormalisation scheme (or scale) variation.
Our results allow for a deeper understanding of the perturbative behaviour of integrated spectral function moments and provide theoretical  support for  low-$Q^2$ $\alpha_s$ determinations.
\end{abstract}

\thispagestyle{empty}

\vspace*{0.0cm}
\tableofcontents
\newpage
\setcounter{page}{1}

\section{Introduction}

Extractions of the strong coupling, $\alpha_s$, at lower energies can be
very precise due to increased sensitivity to the higher-order corrections, as long
as the non-perturbative contributions are under good control.
The prominent example of this type of $\alpha_s$ determination
is the extraction from
inclusive hadronic decays of the $\tau$ lepton, which have been used since
the 90s as a reliable source of information about
QCD dynamics~\cite{Braaten:1988hc,Braaten:1991qm}. Although the decay rate receives a
non-negligible contribution from non-perturbative effects, it is largely
dominated by perturbative QCD, which renders feasible a competitive extraction
of the strong coupling~\cite{Salam:2017qdl,Boito:2014sta,Davier:2013sfa,Pich:2016bdg}. 
Recently, a  similar  $\alpha_s$ determination
 was introduced~\cite{Boito:2018yvl} making use of a compilation of data for $e^+e^-\to {\rm
(hadrons)}$ below the charm threshold~\cite{Keshavarzi:2018mgv}. An attractive feature of this new analysis is that
the systematics is under very good control, although the error due to the data is
still somewhat large.\footnote{This new type of $\alpha_s$ extraction
has been recently included in the 2019 update of the PDG world average, under the ``low-$Q^2$ category"~\cite{Tanabashi:2018oca}.}

Both analyses rely on finite energy sum rules (FESR) where, on the experimental side, one
has weighted integrals of the
experimentally accessible hadronic spectral functions. Exploiting the analyticity properties of the
quark-current correlators one is able to express the theoretical counterpart of the sum rules as an
integral in a closed contour on the complex plane of the variable $s$
--- which represents the invariant mass of the final-state hadrons ---
thereby circumventing the breakdown of perturbative QCD at low
energies.  In this framework, the perturbative contribution
is obtained from the complex integration of the Adler function in the chiral limit, which nowadays is
exactly known up to $\alpha_s^4$~\cite{Baikov:2008jh,Herzog:2017dtz}. When performing this
integration, one must adopt a procedure to set the renormalisation
scale.  The two most widely used ones are Fixed Order Perturbation
Theory (FOPT)~\cite{Beneke:2008ad}, in which the scale is kept fixed, and Contour Improved
Perturbation Theory (CIPT)~\cite{Pivovarov:1991rh,LeDiberder:1992zhd}, where the scale
varies along the contour resumming the running of the coupling. The
procedures lead to different series and to values of $\alpha_s$ that
are different. This difference remains one of the dominant
uncertainties in the $\alpha_s$ extraction from $\tau$ decays~\cite{Boito:2014sta,Pich:2016bdg}. In the case
of $e^+e^-\to {\rm (hadrons)}$, the difference is significantly smaller, but still
non-negligible~\cite{Boito:2018yvl}.

In the discussion of perturbative expansions in QCD one must take into
account a basic but important fact: the perturbative series are divergent and,
at best, they are asymptotic expansions --- as discovered by Dyson in the
context of QED in 1952~\cite{Dyson:1952tj}.
The series is better
understood in terms of its Borel transform, which suppresses the
factorial growth of the perturbative coefficients and allows for
an understanding of the higher-order behaviour  in terms
of singularities along the real axis in the Borel plane. These singularities
are the renormalons of perturbation theory~\cite{Beneke:1998ui}.

An optimal use of an asymptotic series of this type  can be achieved (most often) by
truncating it at the smallest term~\cite{Boyd1999}. In this procedure, the
error one makes is parametrically of the form $e^{-p/\alpha}$ where
$p>0$ is a constant and $\alpha$ the expansion parameter. In QCD, the
expansion parameter, $\alpha_s(Q^2)$, runs logarithmically which
implies that the truncation error is $\sim \left(\Lambda^2_{\rm
  QCD}/Q^2 \right)^p$, where $Q^2$ is the Euclidian momentum. These
power corrections are a necessary feature of perturbative QCD and are,
of course, related to the higher-dimension terms in the Operator Product Expansion (OPE).
In the Borel plane, their manifestation is the appearance of renormalon
singularities  along the real axis at specific locations related to their dimensionality.

In realistic $\alpha_s$ analyses the non-perturbative contributions
must be taken into account. These include the OPE condensates as well
as duality violations (DVs) which are due to resonances and are not encoded
in the OPE expansion~\cite{Shifman:2000jv,Cata:2005zj,Peris:2016jah,Boito:2017cnp} .
In order to extract from the data $\alpha_s$ and the non-perturbative
parameters in a self-consistent way, i.e. without relying on external
information, one resorts to the use of several (pseudo) observables.
Those are built using the fact that any analytic weight function gives
rise to a valid FESR, with an experimental side that can be computed
from the empirical spectral functions and a theoretical counterpart
that can be obtained from the integral along the complex contour.

The main guiding principle behind the judicious choice of weight
functions that enter a given analysis has been, for a long time, the
suppression of non-perturbative contributions. The different analyses of hadronic
tau decay data can be divided into two categories. In one of the analysis
strategies, one strongly suppresses the poorly known higher-order OPE
condensates~\cite{Boito:2014sta,Boito:2018yvl}. In this case,
duality violations are larger and one must include them; this is done relying on
a parametrization that can be connected with fundamental properties of QCD~\cite{Boito:2017cnp}.
In the other, only moments that suppress duality
violations are used~\cite{Davier:2013sfa,Pich:2016bdg}.
 The price to pay
in this case is the contamination of the results by the neglected higher-order OPE condensates~\cite{Boito:2016oam}.
Apart from issues related to non-perturbative contaminations, since the work of  Ref.~\cite{Beneke:2012vb},  it is known that the different weight functions lead to
 distinct perturbative series that are not equally well behaved. Some of those used in the literature~\cite{Davier:2013sfa,Pich:2016bdg}  have a poor perturbative convergence and are therefore not the ideal
choice in precise $\alpha_s$ analyses.

The main purpose of this work is to understand the perturbative behaviour of the different integrated spectral function moments  at intermediate and
high orders by studying the renormalon singularities appearing in their Borel transform.
  The perturbative behaviour of the different moments is
intricate, in fact, each of the moments is a different asymptotic
expansion with different renormalon contributions and conclusions about their
perturbative behaviour have to be drawn almost case by case.  As is customary, we will use the large-$\beta_0$ limit of QCD as a
guide. In this limit, all renormalon singularities are double poles, with the exception of the leading IR singularity, which is simple. A number of facts can be established. First, the  Borel transform of polynomial moments of the Adler function
  is always less singular than the Borel transform of the Adler
  function itself. An infinite number of renormalon poles become simple poles.
Second, the renormalon poles corresponding to the OPE condensate(s) to
  which the moment is maximally sensitive are not reduced (or cancelled).
What we mean by ``maximally sensitive'' will become clearer in the
remainder, but these two facts are enough to draw interesting
conclusions about the behaviour of the different moments and help
explaining the instabilities (or ``run-away behaviour'') identified in Ref.~\cite{Beneke:2012vb}.
We will show that the leading IR renormalon is largely responsible for the unstable  
behaviour of moments that are highly sensitive to the gluon condensate. We also show that an absence of the leading IR pole and  partial cancellations of the renormalon singularities are behind the good perturbative behaviour of some of the moments.

Turning to QCD, the situation is more complicated, mainly because the renormalon poles
become branch points. The Borel transform has superimposed branch cuts. 
We will show that most of the difficulties in QCD can be circumvented by a convenient scheme transformation, to the so-called $C$ scheme~\cite{Boito:2016pwf}, together
with the use of a modified Borel transform introduced in Ref.~\cite{Brown:1992pk}.\footnote{The use of modified Borel transforms in combination with the $C$ scheme in similar contexts has been suggested by M. Jamin and S. Peris~\cite{communication}.} 
In this framework, the Borel transform of the moments can be calculated exactly in terms of the Borel transform of the Adler function --- which is one of the main results of this paper, Eq.~(\ref{eq:modBoreldelta}). The parallel with the large-$\beta_0$ limit is apparent and the results are formally identical. In fact, we show that the leading IR singularity is also a simple pole in this case. The enhancement and suppression of renormalon singularities identified in large-$\beta_0$ is, therefore, also present in QCD which
explains the similarity between the perturbative behaviour of moments in the two cases. We then study the behaviour of a few emblematic moments in QCD, using a recent reconstruction of the higher-order terms based on Pad\'e approximants~\cite{Boito:2018rwt}. Finally, we show how to optimize the truncation of the moments with good perturbative behaviour in the spirit of an asymptotic series exploiting scheme transformations. The procedure we employ has been suggested for the $\tau$ hadronic width in Ref.~\cite{Boito:2016pwf} but had never been investigated systematically for different integrated moments.

This work is organised as follows. In Sec.~\ref{theory}, we present
the theoretical framework. In Sec.~\ref{renormalons}, we discuss
the renormalon content of polynomial moments and their phenomenological consequences, both  in large-$\beta_0$ and in QCD. In Sec.~\ref{optimization}, we discuss the optimized truncation
of the moments with good perturbative behaviour through scheme transformations. In Sec.~\ref{conclusions}, we present our conclusions. Finally, in App.~\ref{betafunction} we
present our conventions for the QCD $\beta$ function;  App.~\ref{borelint} contains further details about the Borel integral of the moments discussed in this work.

\section{Theoretical framework}
\label{theory}

In the low-$Q^2$ $\alpha_s$ determinations from hadronic $\tau$ decays and from $\epem$
one uses FESRs constructed from integrated moments of the experimental hadronic spectral functions.
In the case of $\epem$, one has access to the electromagnetic vacuum polarization spectral function, which mixes isospin 0 and 1. Below the charm threshold one can safely work in the chiral limit, apart from the inclusion of perturbative corrections arising from the strange-quark mass. In hadronic $\tau$ decays, the decay width of the $\tau$ lepton into hadrons normalized to the
decay width of $\tau\to \nu_\tau e^- \bar\nu_e$ can be separated experimentally
into three distinct components: the vector and axial vector, $R_{\tau, V/A}$,
arising from the $(\bar ud)$-quark current, and the contributions with net
strangeness, intermediated by the $(\bar us)$-quark current. In the
extractions of the strong coupling $\alpha_s$, the focus is on the non-strange
contributions since they have a smaller contamination from non-perturbative
effects and the quark masses in this case can safely be neglected. Since 
the FESRs we discuss here, and in particular the choice of moments, were primarily introduced in the context of $\tau$ decays, we will present them in this context. The translation to $\epem$ is straightforward and the perturbative contribution, in particular, is essentially identical~\cite{Boito:2018yvl}.

We define a generalized observable  $R_{\tau, V/A}^{(w_i)}(s_0)$ that can be written as a weighted integral over the
experimentally accessible spectral functions as
\beq
R_{\tau, V/A}^{(w_i)}(s_0) = 12\pi S_{\rm EW} |V_{ud}|^2 \int_0^1 dx\, w_i(x) \left[{\rm Im}\Pi_{V/A}^{(1+0)}(xs_0)+\frac{2x}{1+2x}{\rm Im}\Pi_{V/A}^{(0)}(xs_0) \right],\label{eq:expmoment}
\eeq
where $w_i(x)$ is any analytic weight function and $x=s/s_0$. The correlators $\Pi^{(1)}_{V/A}$ and $\Pi^{(0)}_{V/A}$
are the transverse and longitudinal parts of
\beq
\Pi_{V/A}^{\mu \nu}(p) \equiv i \int dx \,e^{i p x}\, \langle \Omega | T \{ J_
{V/A}^{\mu}(x) J_{V/A}^{\nu}(0)^{\dagger} \} | \Omega \rangle,\label{eq:Corr}
\eeq
formed from the quark currents $J_{V/(A)}^\mu = (\bar u \gamma^\mu(\gamma_5) d)(x)$; we define $\Pi^{(1+0)}_{V/A}=\Pi^{(1)}_{V/A}+\Pi^{(0)}_{V/A}$. 
Setting $s_0=m_\tau^2$ in Eq.~(\ref{eq:expmoment}) and  with the particular choice of weight function
\beq
w_\tau(x) = (1-x)^2(1+2x)
\eeq
dictated by kinematics  we have $R_{\tau,{V/A}}\equiv R_{\tau,V/A}^{(w_\tau)}(m_\tau^2)$, the hadronic decay width
normalized to the decay width of $\tau^- \to \nu_\tau e^- \bar{\nu}_e $.

 In precise extractions of $\alpha_s$ from
$\tau$ decays it has become customary to exploit other analytic weight functions,
conveniently chosen in order to suppress or enhance the different contributions
to the decay rate. The generalized observable  $R_{\tau,V/A}^{(w_i)}(s_0)$ can be decomposed as
\beq
R_{\tau,V/A}^{(w_i)}(s_0) = \frac{N_c}{2}S_{\rm EW} |V_{ud}|^2 \left[\delta_{w_i}^{\rm tree}
+ \delta_{w_i}^{(0)}(s_0) + \sum\limits_{D\geq2}\delta_{w_i,V/A}^{(D)}(s_0)
+ \delta^{\rm DV}_{w_i,V/A}(s_0)\right],\label{eq:Rtaudecomposition}
\eeq
where $N_c$ is the number of colours, $S_{\rm EW}$ is an electroweak correction,
and $V_{ud}$ is the quark-mixing matrix element. The perturbative terms are represented
by $\delta_{w_i}^{\rm tree}$ and $\delta_{w_i}^{(0)}(s_0)$, where the former
corresponds to the partonic result, while the latter encode the $\alpha_s$ corrections
computed in the chiral limit. The OPE corrections of dimension $D$ are collected
in the terms $\delta_{w_i,V/A}^{(D)}(s_0) $ and, finally, duality violation
corrections are given by  $\delta^{\rm DV}_{w_i,V/A}(s_0)$.

The leading contribution to $R_{\tau,V/A}^{(w_i)}(s_0)$ stems from perturbative QCD. It is
obtained from the perturbative expansion of the correlators of Eq.~(\ref{eq:Corr}), which 
is the dimension zero term in the OPE expansion that can be written, for $\Pi^{(1+0)}(s)$, as
\beq
\Pi_{\rm OPE}^{(1+0)} = \sum_{D=0,2,4...}^\infty \frac{C_{D}(s)}{(-s)^{D/2}},\label{PiOPE}
\eeq
where the sum is done over  all the contributions from gauge invariant operators of dimension $D$.
The case $D=0$ is the perturbative part and $D=2$ are small mass corrections. The first non-perturbative contribution starts at $D=4$ and is
dominated by the gluon condensate. The  $s$ dependence in the Wilson coefficients
$C_D(s)$  arise from the logarithms in their perturbative description and is higher-order
in $\alpha_s$. In the case of the gluon condensate the leading logarithm  is known but,
to an excellent approximation, the coefficient $C_4(s)$ can be treated as a
constant~\cite{Boito:2011qt}. Little is known for the logarithms in the higher-dimension
condensates, but it is customary, based on the experience with $D=4$, to neglect their
$s$ dependence as well and treat all $C_D$ as effective coefficients with no $s$ dependence.

The theoretical treatment of the observables $R_{\tau, V/A}^{(w_i)}$ is done in the
framework of FESRs, relating the experimental results to counter-clockwise contour
integrals along the circle $|s|=s_0$ in the complex plane of the variable $s$.
To eliminate the conventions related to renormalisation it is convenient to work
with the Adler function
\beq
D(s) = -s \frac{d}{ds}\Pi^{(1+0)}(s).
\eeq
In terms of the Adler function, the perturbative correction of Eq.~(\ref{eq:Rtaudecomposition}) can be written as~\cite{Beneke:2008ad}
\beq
 \delta^{(0)}_{w_i} = \frac{1}{2\pi i}  \oint\displaylimits_{|x|=1}\frac{dx}{x} W_i(x)
 \widehat D_{\rm pert}(s_0 x),\label{eq:delta0}
\eeq
where $W_i(x)=2\int_x^1 dz \ w_i(z)$ is the weight function. The reduced
Adler function, $\widehat D$, which intervenes in Eq.~(\ref{eq:delta0}), is
defined in order to separate the partonic contribution
\beq
1+ \widehat D(Q^2) =  \frac{12\pi^2}{N_c} D(Q^2),
\eeq
where $Q^2 \equiv -s $. Accordingly, the  perturbative expansion of the function $\widehat D$  starts at
order $\alpha_s$ and can be written as
\beq
\widehat D_{\rm pert}(s) =  \sum\limits_{n = 1}^{\infty}{a^{n}_{\mu}}
\sum\limits_{k = 1}^{n+1 }{k\, c_{n,k}\left[\log\left(-s/\mu^2 \right)\right]^{k-1}},\label{eq:AdlerExp}
\eeq
where $a_\mu=\alpha_s(\mu)/\pi$. The only independent coefficients in this
expansion are the $c_{n,1}$; all the others can be written with the used of
Renormalisation Group (RG) equations in terms of the $c_{n,1}$ and $\beta$-function
coefficients.  At present, the coefficients of the expansion are known up to $c_{4,1}$ (five loops)~\cite{Baikov:2008jh,Herzog:2017dtz}.
Resumming the logarithms with the choice $\mu^2=-s$ the result is (for $n_f=3$)
\beq
\widehat D_{\rm pert}(Q^{2}) =  \sum_{n=1}^\infty c_{n,1} a_Q^{n} =
a_Q + 1.640 \, a_Q^2 +6.371\, a_Q^3+ 49.08\, a_Q^4+\cdots,\label{eq:DCIPT}
\eeq
from which the known independent coefficients can be read off. (Henceforth we will often omit the subscript ``pert'' in perturbative quantities.)

The perturbative series of Eq.~(\ref{eq:AdlerExp}) is divergent. It
is assumed that it must be an asymptotic series~\cite{Beneke:1998ui} to the
true (unknown) value of the function being expanded. The divergence stems
from the factorial growth of the $c_{n,1}$ coefficients at large order and
it is, therefore, convenient to work with the Borel-Laplace transform of the series
\beq
B[\widehat D](t) \equiv \sum\limits_{n=0}^{\infty}{r_{n} \frac{t^{n}}{n !}},
\label{BorelDef}
\eeq
which has a finite radius of convergence and where $r_n = c_{n+1,1}/\pi^{n+1}$. The original expansion is then, by
construction, the asymptotic series to the inverse Borel transform (the usual
Laplace transform) given by
\beq
\widehat D (\alpha) \equiv \int\limits_{0}^{\infty}{dt \text{e}^{-t/\alpha}B[\widehat R](t)} \label{BorelInt}.
\eeq
On the assumption that  the integral exists, the last equation defines
unambiguously the Borel sum of the asymptotic series. However, the
divergence of the original series is related to singularities in the
$t$ variable known as renormalons. They appear at both positive and
negative integer values of the variable $u=\frac{\beta_1 t}{2\pi}$
(with the exception of $u=1$). In particular,  the IR renormalons, that
lie on the positive real axis,  obstruct the integration in the Borel sum.
A prescription to circumvent these poles becomes necessary, which entails
an ambiguity in the Borel sum of the series. This remaining ambiguity is
expected on general grounds to be cancelled by corresponding ambiguities
in the power corrections of the OPE. At large orders, the UV pole at $u=-1$,
being the closest to the origin,  dominates the behaviour of the series.
The coefficients of the series are, therefore, expected to diverge with
sign alternation at sufficiently high orders.

The calculation of the perturbative contribution to FESR
observables requires that one performs the integral of Eq.~(\ref{eq:delta0}).
A prescription for the renormalisation scale $\mu$ --- which enters through
the logarithms of Eq.~(\ref{eq:AdlerExp}) --- must be adopted in the process.
In the procedure known as Contour-Improved Perturbation Theory (CIPT)~\cite{Pivovarov:1991rh,LeDiberder:1992zhd} a
running scale $\mu^2=Q^2$ is adopted and the running of $\alpha_s$  is
resummed along the contour with the QCD beta function. With this procedure the
perturbative contribution is cast as
\beq
\delta^{(0)}_{{\rm CI},w_i} = \sum_{n=1}^{\infty} c_{n,1}J^{(n)}_{{\rm CI},w_i}(s_0), \,\,\,\,\,
\mbox{with} \,\,\, \,\,J^{(n)}_{{\rm CI},w_i}(s_0)=  \frac{1}{2\pi i}\oint\displaylimits_{|x|=1}
\frac{dx}{x}W_i(x)a^{n}(-s_0 x).
\eeq

A strict fixed order prescription, known as Fixed Order Perturbation Theory (FOPT)
corresponds to the choice of  a fixed scale $\mu=s_0$. The coupling can then be
taken outside the integrals which are now performed over the logarithms  of
Eq.~(\ref{eq:AdlerExp}) as
\beq
\delta^{(0)}_{{\rm FO},w_i} = 	\sum\limits_{n=1}^{\infty}{a_{s_0}^{n}}
\sum\limits_{k=1}^{n}{k c_{n,k}} J_{{\rm FO},w_i}^{(k-1)},\,\,\,\,\, \mbox{with} \,\,\,\,\,
J_{{\rm FO},w_i}^{(n)} \equiv \frac{1}{2 \pi i}\oint\limits_{|x|=1}{\frac{dx}{x}W_i(x)\ln^{n}(-x)}.\label{FOPTdef}
\eeq
The FOPT series can be written as an expansion in the coupling as
\beq
\delta_{{\rm FO},w_i}^{(0)} = \sum_{n=1}^\infty  d_n^{(w_i)} a_Q^n,
\eeq
where the coefficients now depend on the choice of weight function.

The chosen prescription for the renormalisation group improvement of the
series affects, in practice, the precise extraction of the strong coupling
from hadronic $\tau$ decays. It remains, as of today, one of the main
sources of theoretical uncertainty in these $\alpha_s$ determinations~\cite{Pich:2016bdg,Davier:2013sfa,Boito:2014sta,Boito:2018yvl}.
The two prescriptions define two different asymptotic series with rather different
behaviours. Inevitably, the analysis of the  reliability  of the two procedures
requires  knowledge about higher orders of the series. In particular, some of
the arguments often put forward in favour of CIPT --- in an attempt to leave
aside the issue with the higher orders ---  mention a ``radius of convergence"~\cite{Braaten:1991qm,Pich:2013lsa}, a
notion that  contradicts the fact that the series are both asymptotic.

Here we employ the estimate for the higher-order coefficients of the series obtained from a careful and systematic use of Pad\'e approximants~\cite{Boito:2018rwt}. The results of Ref.~\cite{Boito:2018rwt} are model independent and corroborate to a large extent the results obtained in the context of renormalon models, in which the series is modelled by a small number of dominant renormalon singularities employing the available knowledge about their nature~\cite{Beneke:2008ad,Beneke:2012vb,Cvetic:2018qxs}, as well as those obtained from conformal mappings that make use of the location of renormalon sigularities~\cite{Caprini:2019kwp}.
We will also exploit scheme variations as a method to improve convergence of the perturbative series and discuss their usefulness in realistic extractions of $\alpha_s$ from hadronic $\tau$ decays.

\section{Renormalons in spectral function moments}
\label{renormalons}

Several  moments of the spectral functions have been used in low-$Q^2$ $\alpha_s$
determinations from hadronic $\tau$ data and $\epem$~\cite{Boito:2011qt,Davier:2013sfa,Boito:2014sta,Boito:2016oam,Pich:2016bdg}.
Since the FESR requires the weight function to be analytic it is customary
to employ  polynomials,
which we denote
in terms of their expansions in monomials as
\beq
w_i(x) = \sum_{k=0}b^{(w_i)}_k\,x^k.
\eeq
Of particular importance are the weight functions that are
``pinched", i.e. weight functions that are zero at $x=1$,
and  that have $b_0^{(w_i)}=1$ 
such as the kinematic moment
\beq
w_\tau = (1-x)^2(1+2x) = 1-3x^2+2x^3\label{wtau}.
\eeq
Another important class of weight functions identified in~\cite{Beneke:2012vb} are those that contain the linear term in $x$. We will discuss these two classes of moments in detail below.

The series for $\delta^{(0)}_{w_i}$ inherits the divergence of the Adler
function expansion and accordingly is also amenable to a treatment in terms
of its Borel transform.  However, the renormalon content of the Borel transformed
$\delta^{(0)}_{w_i}$ is different from the Adler function counterpart, as we discuss
in the remainder of the section.

\subsection[Results in  \texorpdfstring{large-$\beta_0$}{large-beta0}]{Results in \boldmath \texorpdfstring{large-$\beta_0$}{large-beta0}}
We start investigating the renormalons in $\delta^{(0)}_{w_i}$ in the
large-$\beta_0$ limit of QCD~\cite{Beneke:1998ui}. These results are obtained by
first considering a large number of fermion flavours, $N_f$, but keeping
$N_f\alpha_s$ constant. The $q\bar q$ bubble corrections to the gluon
propagator are order one in this power counting and must be summed to all orders. This dressed gluon
propagator is used to obtain all the leading $N_f$ corrections, at every
$\alpha_s$ order, to a given observable. In the end, $N_f$ is replaced by
the leading $\beta$  function coefficient, effectively incorporating a set of
non-abelian contributions~\cite{Beneke:1994qe}. Accordingly, the
$\alpha_s$ evolution is performed at one-loop.

In this limit, the Borel transformed Adler function is known to all orders
in perturbation theory and it can be written in a compact form
as~\cite{Beneke:1998ui,Beneke:1992ch,Broadhurst:1992si}
\beq
B[\widehat D]= \frac{32}{3\pi} \frac{\text{e}^{(C+5/3)u}}{(2-u)} \sum\limits_{k=2}^{\infty} \frac{(-1)^k k}{[k^2-(1-u)^2]^2},
\label{eq:BorelAdlerLb0}
\eeq
where $C$ is a parameter which depends on the renormalisation scheme.
For $C=0$ we have $\overline{\text{MS}}$. This result displays explicitly
the renormalon poles. They are all double poles with the exception of the
leading IR pole at $u=2$, which is simple.  The IR poles are particularly
important in the subsequent discussion, and in particular their connection
to OPE condensates. Each of the IR poles that appear at a given position $u=p$
in the Borel transform of the Adler function can be mapped to the existence of
contributions  of dimension $D=2p$ in the OPE~\cite{Beneke:1998ui}. This
explains, for example, the absence of a pole at $u=1$ since there is no gauge-invariant $D=2$
condensate in the OPE.  This non-trivial connection between perturbative
and non-perturbative physics will also be manifest in the Borel transform of
$\delta_{w_i}^{(0)}$.

Using this result, the Borel transform of $\delta_{w_i}^{(0)}$ can be obtained
from  Eq.~(\ref{eq:delta0}) employing
the Borel integral representation of the Adler function, Eq.~(\ref{BorelInt}).
One can then write
\beq
\delta^{(0)}_{w_i} =\frac{1}{2\pi}\int_0^{2\pi}d\phi\, W_i(e^{i\phi}) \int_0^\infty dt e^{-t/\alpha_s(-s_0e^{i\phi})}B[\widehat{D}](t),\label{eq:delta0intrepre}
\eeq
where we performed the change of variables $x=e^{i \phi}$. In the large-$\beta_0$ limit, the $\beta$ function is truncated at its first term\footnote{For our conventions regarding the QCD $\beta$ function we refer to App.~\ref{betafunction}}
\beq
\frac{1}{\alpha_s(Q^2)}=\frac{\beta_1}{2\pi}\ln\left(\frac{Q^2}{\Lambda^2}\right).\label{eq:alphasrunning1loop}
\eeq
The exponential in Eq.~(\ref{eq:delta0intrepre}) can be written as
\beq
e^{-t/\alpha_s(-s_0x)}=e^{-t/\alpha_s(s_0)}e^{-iu(\phi-\pi)}.
\eeq
Inverting the order of integration and using Eq.~(\ref{BorelInt}) one can read off the
Borel transform of $\delta^{(0)}_{w_i}$
\beq
B[\delta^{(0)}_{w_i}] = \left[\frac{1}{2\pi}\int_0^{2\pi}d\phi\, W_i(e^{i\phi})e^{-i u (\phi-\pi)}    \right] B[\widehat D](u).
\label{BTdLb1}
\eeq
The prefactor of Eq.~(\ref{BTdLb1}) can
be obtained analytically for polynomial weight functions. For the monomial
$w_i=x^n$ one finds\footnote{This result has been  used in Refs.~\cite{Boito:2018rwt,Caprini:2019kwp}.}
\beq
B[\delta^{(0)}_{x^n}] = \frac{2}{1+n-u}\frac{\sin (\pi u)}{\pi u} B[\widehat D](u).
\label{BTdLb2}
\eeq
One immediately sees that the $\sin(\pi u)$ reduces an infinite number of
UV and IR double poles in $B[\widehat D](u)$  to simple poles. In this
sense, one can say that $B[\delta^{(0)}_{w_i}]$ is significantly less
singular than the Adler function counterpart, a fact that has been exploited
in Ref.~\cite{Boito:2018rwt}.

The prefactor of Eq.~(\ref{BTdLb2}) is also highly non-trivial.  It cancels
the zero at $u=1+n$ in $\sin(\pi u)$, which means that the pole at $u=1+n$
of $B[\widehat D](u)$ remains double (or single, in the case of $u=2$). This
is clearly not a coincidence and is related to the non-perturbative contributions
to $R_{\tau,V/A}$. To expose this connection, consider the contribution of $D\geq 4$
in the OPE expansion, Eq.~(\ref{PiOPE}), to  $R_{\tau,V/A}$ which can be cast as
\beq
\delta_{w_i,V/A}^{(D)}=\frac{6\pi i}{(-s_0)^{D/2}} \oint dx \frac{w_i(x)}{x^{D/2}} C_D(xs_0).
\eeq
For a monomial $w_i(x)=x^n$ --- and to the extent that the $s$ dependence of the
coefficients $C_D$ can be neglected, as discussed previously --- this reduces to
\beq
\delta_{x{^n},V/A}^{(D)}=\frac{6\pi i}{(-s_0)^{D/2}} C_D \oint dx \frac{1}{x^{-n+D/2}}.\label{OPEdeltaD}
\eeq
For positive integer values of $n$, the integral in the last equation is only
non-vanishing for $-n+D/2=1$. Therefore, as is well known, under these assumptions, for $w_i=x^n$ the  only  contribution
comes from the condensates with $D=2(n+1)$ which, in turn, is related to the
pole in the Borel transform of the Adler function at $u=n+1$.

It becomes apparent that the prefactor of Eq.~(\ref{BTdLb2}) is not accidental: the pole in
$B[\widehat D](u)$ that corresponds to the condensate that contributes maximally to moments
of $w=x^n$ is {\it not} cancelled by the prefactor of Eq.~(\ref{BTdLb2}). For monomials $x^n$ with $n\geq0$ three cases can be distinguished:
\begin{itemize}
		\item If $n=0$ all poles become simple poles, since there is no contribution
	from OPE condensates, apart from the $\alpha_s$-suppressed terms, under the assumptions of Eq.~(\ref{OPEdeltaD}). In particular, the pole at
	$u=2$ which was simple is exactly cancelled and the function is regular at $u=2$.

\item If $n=1$, the dominant contribution from the OPE is the one from $D=4$. The pole
at $u=2$ related to this OPE contribution is not canceled and all other IR and UV poles
become simple poles. This is a distinct situation because it is the only case where
$B[\delta^{(0)}_{x^n}]$ is singular at $u=2$, in all other cases the  leading  IR singularity
is located at $u=3$.

\item  Finally, if $n\geq 2$, all IR poles for $u>2$ become simple poles, with the exception of the
pole at $u=n+1$, which remains double and is now the  only double pole in
$B[\delta^{(0)}_{x^n}]$ --- all others are reduced to simple poles by the zeros of
$\sin(\pi u)$. In this case, the pole at $u=2$ that corresponds to the contributions
due to the gluon condensate is again exactly cancelled by $\sin(\pi u)$ and the function is
analytic at $u=2$.
\end{itemize}

\begin{table}[!t]
	\begin{center}
		{		\caption{Residues for the dominant poles in the Borel transform of $\delta^{(0)}_{w_i}$ for the first six monomials. Boxed numbers refer to residues of double poles, all other poles are simple poles.}
			\begin{tabular}{cccccccccc}
				\toprule
				$w(z)$	&  $u=-2$ & $u=-1$ & $u=1$ & $u=2$ & $u=3$ & $u=4$ & $u=5$ \\
				\midrule
				$1$    &  $8.411\times10^{-4}$  &  $2.672\times10^{-2}$  &  $0$  &  $0$      &  $-21.00$  &  $-18.53$  &  $-29.43$  \\
				$z$    &  $6.309\times10^{-4}$  &  $1.781\times10^{-2}$  &  $0$  &  $17.85$  &  $-41.99$  &  $-27.79$  &  $-39.24$  \\
				$z^2$  &  $5.047\times10^{-4}$  &  $1.336\times10^{-2}$  &  $0$  &  $0$      &  $\boxed{6.999}$   &  $-55.58$  &  $-58.86$  \\
				$z^3$  &  $4.206\times10^{-4}$  &  $1.069\times10^{-2}$  &  $0$  &  $0$      &  $41.99$   &  $\boxed{-32.42}$  &  $-117.7$  \\
				$z^4$  &  $3.605\times10^{-4}$  &  $8.907\times10^{-3}$  &  $0$  &  $0$      &  $21.00$   &  $55.58$   &  $\boxed{-104.0}$  \\
				$z^5$  &  $3.154\times10^{-4}$  &  $7.634\times10^{-3}$  &  $0$  &  $0$      &  $14.00$   &  $27.79$   &  $117.7$   \\
				\bottomrule
			\end{tabular}
			\label{tab:w_monomial}
		}
	\end{center}
\end{table}

In Tab.~\ref{tab:w_monomial}, we show the residues of the dominant UV and IR poles for the first six
monomials, which are the building blocks for most of the moments used in the literature. Residues of the double poles are shown as boxed numbers. The results of this table can be used to understand
a few features of specific cases. For example, in the Borel transform of the kinematic moment,
Eq.~\eqref{wtau}, a partial cancellation of the leading UV renormalon is manifest: its residue is reduced
by a factor of 3.3. The perturbative series associated with this moment is expected to display a more
tamed behaviour, with the asymptotic nature setting in later.

We  are now in a position to reassess some of the findings of Ref.~\cite{Beneke:2012vb} in the light of these
results.
One of the main observations of Ref.~\cite{Beneke:2012vb}  is that the perturbative series for moments of weight
functions that
contain the monomial $x$ tend to be badly behaved, in the sense that the series never stabilise around
the true value of
the function, it displays what was called a ``run-away behaviour". This can be directly linked to the fact
that the Borel
transform of these moments are the only ones that have the singularity at $u=2$. The contribution of this
renormalon
to the coefficients of the series is fixed sign and it is large at higher orders.

\begin{figure}[!t] 
\begin{center}
{\includegraphics[width=.7\columnwidth,angle=0]{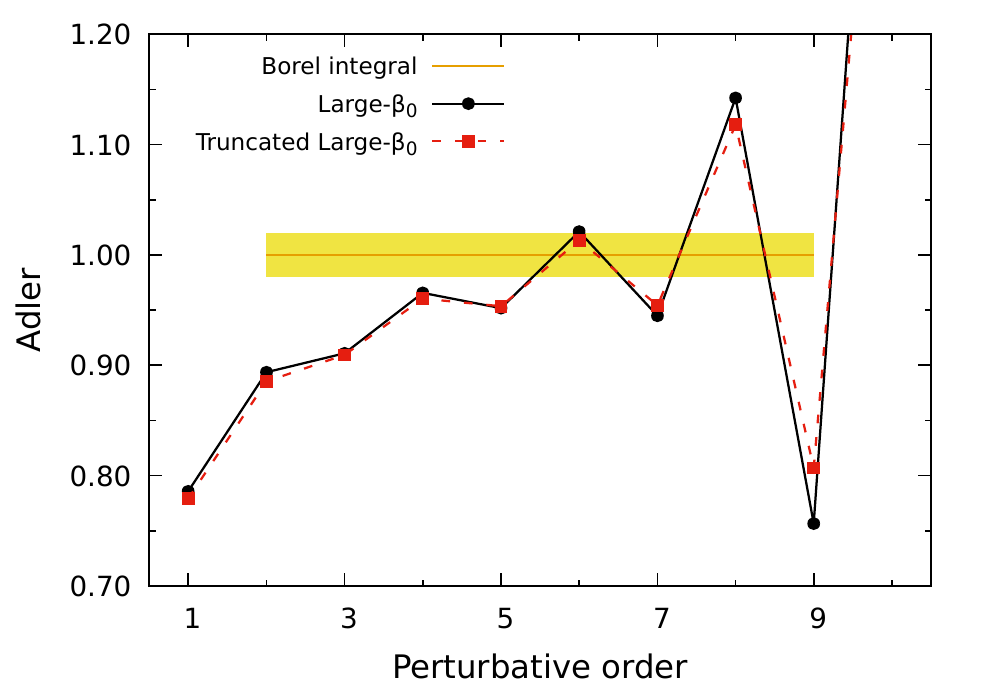}}
\caption{Adler function order by order in $\alpha_s$ in large-$\beta_0$ and in the truncated version that includes only the leading UV and the first two IR renormalon poles. Both results are normalised to the respective Borel integrals. The horizontal band gives the ambiguity arising from the IR poles, in the prescription of Ref.~\cite{Beneke:2008ad}. Here and elsewhere we use $\alpha_s(m_\tau^2)=0.316(10)$~\cite{Patrignani:2016xqp}.}
\label{fig:Adler_Lb}
\end{center}
\end{figure}

In order to establish the correspondence between the leading renormalons and the behaviour of the
perturbative series, we will make use of an even simpler model.
Since the series is dominated by the leading renormalons, we can construct an approximation to
the result in large-$\beta_0$ using only the leading UV pole and the first two IR poles, which
corresponds to truncating the sum in Eq.~\eqref{eq:BorelAdlerLb0} at its first term. We know from the
works of Ref.~\cite{Beneke:2008ad,Beneke:2012vb,Boito:2018rwt} that such a minimalistic model should be largely sufficient to capture the
main features of the full result in large-$\beta_0$. In Fig.~\ref{fig:Adler_Lb} we confirm this
expectation by plotting the results for the Adler function in large-$\beta_0$ and in its truncated
version, normalized to the value of the Borel integral in each case, which removes an overall
normalization effect that is immaterial here (throughout this paper we use $\alpha_s(m_\tau^2)=0.316(10)$~\cite{Patrignani:2016xqp}). In Fig.~\ref{fig:Adler_Lb}, one sees that the results are essentially identical for our purposes. However, the simplicity of this model prevents the study of moments that are maximally sensitive to
condensates with dimension $D \geq 8$, because the corresponding renormalon poles  are not included.

\begin{figure}[!t]
	\begin{center}
		\subfigure[$\delta^{(0)}$, $w(x)=x$, truncated large-$\beta_0$]{\includegraphics[width=.49\columnwidth,angle=0]{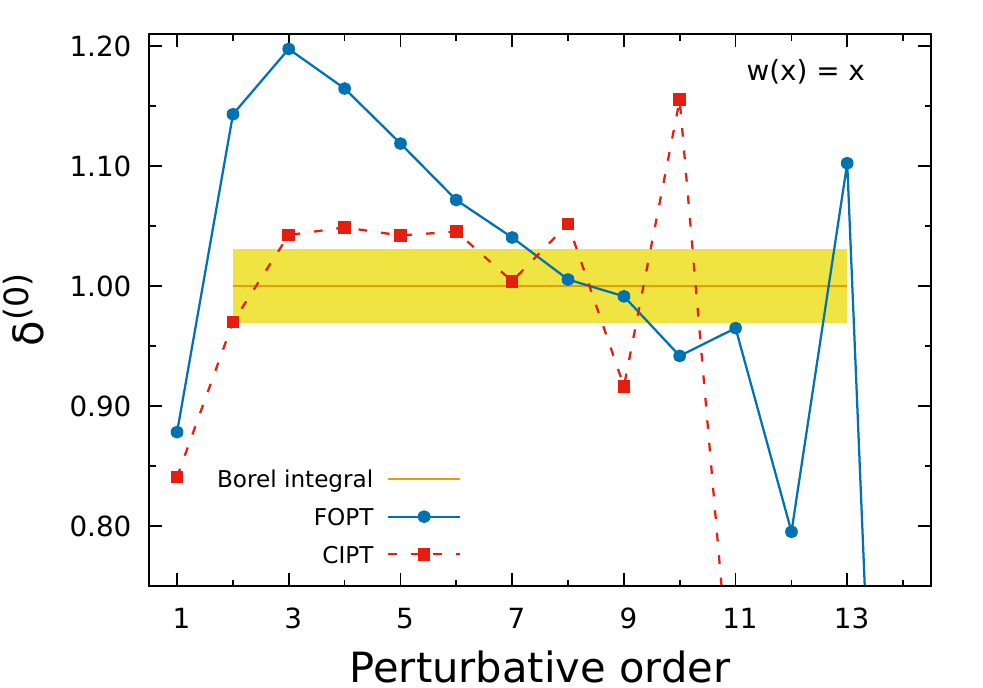}\label{fig:TruncatedLbweqx}}
		\subfigure[$\delta^{(0)}$, $w(x)=1-x^2$, truncated large-$\beta_0$] {\includegraphics[width=.49\columnwidth,angle=0]{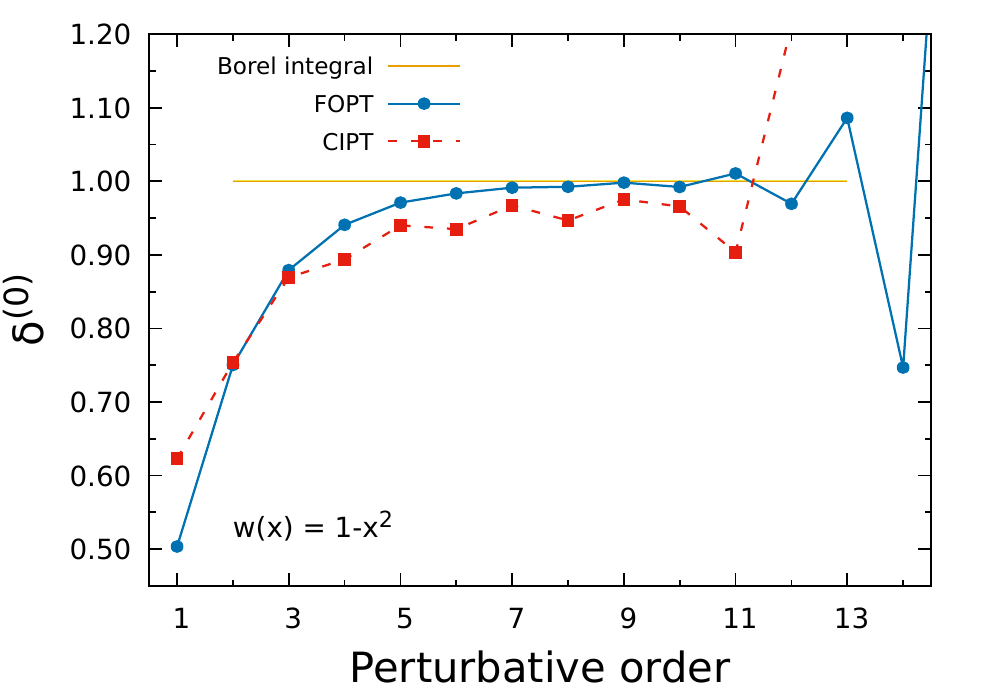}\label{fig:TruncatedLbweq1mx2}}
		\subfigure[Contributions to $\delta^{(0)}_x$ (FOPT).]{\includegraphics[width=.49\columnwidth,angle=0]{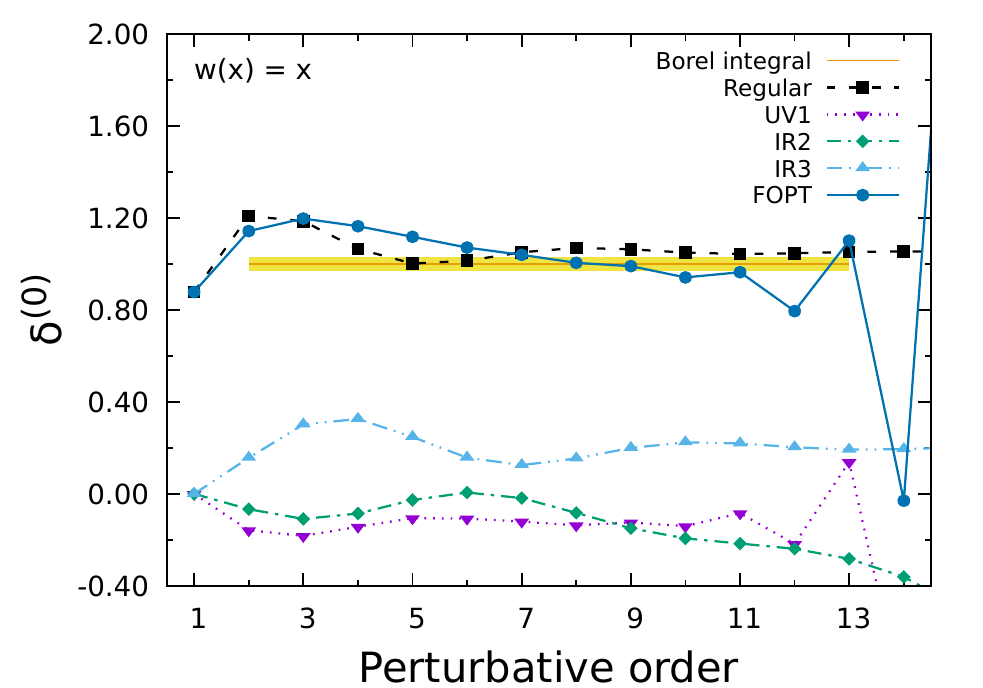}\label{fig:TruncatedLbPolesweqx}}
		\subfigure[Contributions to $\delta^{(0)}_{1-x^2}$ (FOPT).] {\includegraphics[width=.49\columnwidth,angle=0]{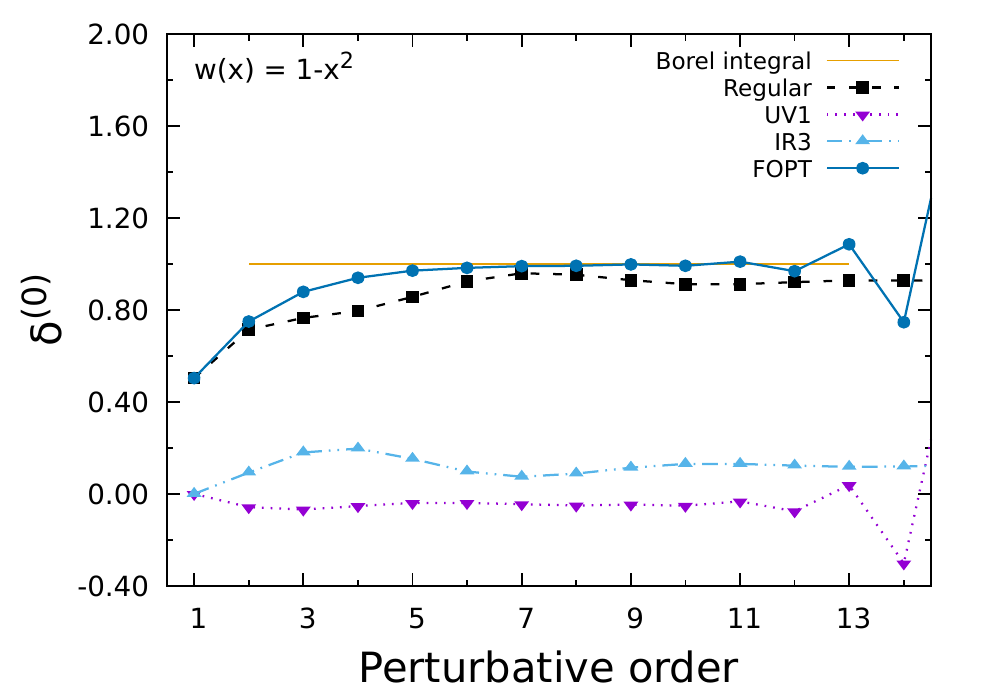}\label{fig:TruncatedLbPolesweq1mx2}}
		\caption{Perturbative series for the moments $w(x)=x$ and $w(x)=1-x^2$ order by order in $\alpha_s$ in the truncated version of the large-$\beta_0$ limit (see text for details) with $s_0=m_\tau^2$ (upper panels).  Breakdown of the contributions from the leading  singularities for the same moments in FOPT (lower panels). Results are normalized to the respective Borel integrals. The horizontal band represents the ambiguity arising from IR renormalon poles, calculated as in Ref.~\cite{Beneke:2008ad}.}
		\label{fig:TruncatedLbw}
	\end{center}
\end{figure}

We start by considering the moment of $w(x)=x$. The perturbative expansion of $\delta^{(0)}_x$ in the
truncated model for FOPT and CIPT are displayed in Fig.~\ref{fig:TruncatedLbweqx}. The FOPT series shows
the ``run-away behaviour'' identified in Ref.~\cite{Beneke:2012vb}. It overshoots the true value, at first, and
later crosses it and runs into the asymptotic regime with almost no stable region. The CIPT series is
better behaved but also overshoots the true value and then runs into the asymptotic behaviour, with sign
alternating coefficients, much earlier than FOPT. To understand this pattern we can use the Borel
transform of $\delta^{(0)}_{x}$ which is rather simple in the truncated model,
\begin{equation}
B[\delta^{(0)}_x]_T (u) = \frac{4 \text{e}^{5u/3}\sin(\pi u)}{27\pi^2}\Big{[} \frac{8}{u} + \frac{8(5u-4)}{3(2-u)^2} -
\frac{(13u+19)}{3(1+u)^2} - \frac{(17 u - 57)}{(3-u)^2} \Big{]}.
\end{equation}
The result exhibits the UV pole at $u=-1$, as well as the IR poles at $u=2$ and $u=3$. All poles are
simple due to the zeros of the $\sin(\pi u)$ in the prefactor. We also note that the Borel transform has
a regular part, which stems from the first term within square brackets (the would-be pole at zero is also
canceled by the prefactor). In Fig.~\ref{fig:TruncatedLbPolesweqx} we show the breakdown of the different contributions to the perturbative series in FOPT. The series is dominated by the regular contribution which initially overshoots the true value. At higher orders, the first IR and UV poles dictate the tendency and the series never stabilizes around the true value. The IR contribution is negative and is responsible for the run-away behaviour, with a superimposed sign alternation from the UV pole.

We now turn to the pinched moments without the term in $x$.
In Fig.~\ref{fig:TruncatedLbweq1mx2}, we show the results for $w=1-x^2$.
 The Borel transform of $\delta^{(0)}_{1-x^2}$ is regular
at $u=2$ and the leading UV pole is partially cancelled, as we can infer from the results of
Tab.~\ref{tab:w_monomial}. This  translates into a smoother series.
Now the FOPT series nicely approaches the true value and remains stable around it for several orders
until eventually entering the asymptotic regime, when the leading UV pole takes over. The result for CIPT, on the other hand, is less accurate (red dashed line in Fig.~\ref{fig:TruncatedLbweq1mx2}). It approaches the Borel sum of the series only when the asymptotic behaviour has already set in.  

Finally, we comment on the results
for $w(x)=1$. This moment lies somewhere in between the two extreme cases we discussed above. It also benefits from being regular at $u=2$ but the partial cancellation of singularities that happens in pinched moments is not present.
In this case, FOPT is  able to approach the result although at the expense of overshooting it for the first four orders or so. We omit the plots in this case  for the sake of brevity (the result in the context of Borel models can be found in Ref.~\cite{Beneke:2012vb}). One should finally remark that, in general, the perturbative series for $\delta^{(0)}_{w_i}$ are  better behaved than the Adler function series, shown in Fig.~\ref{fig:Adler_Lb}. This fact is a consequence of the Borel transform of $\delta^{(0)}_{w_i}$ being significantly  less singular than the Adler function counterpart. The sign alternation in the Adler function starts already at $\mathcal{O}(\alpha_s^5)$ and the perturbative series never stabilises around the Borel sum.
This does not prevent, however, the pinched moments without the linear term from having a very good perturbative behaviour, as exemplified in Fig.~\ref{fig:TruncatedLbweq1mx2}.

\subsubsection{Partial conclusions}
We are in a position to draw a few conclusions from the study of spectral function moments in large-$\beta_0$ and its truncated form:
\begin{itemize}
	\item The Borel transform of $\delta^{(0)}_{w_i}$ for polynomial weight functions is less singular than the Borel transform of the Adler function. The prefactor that appears in $B[\delta^{(0)}_{w_i}]$ cancels an infinite number of poles. Precisely for this reason, the perturbative expansions of $\delta^{(0)}_{w_i}$ are in general smoother than the Adler function counter part.
		\item The pattern of the remaining poles in the Borel transform of $\delta^{(0)}_{w_i}$ can be understood in terms of the contributions from the OPE condensates.
	 The Borel transform has a pole  at $u=2$ if and only if the weight function contains a term proportional to $x$. The behaviour of the perturbative series associated with these moments is qualitatively different and the true value of series is not well approached neither by FOPT nor by CIPT, as already discussed in Ref.~\cite{Beneke:2012vb}.
	 \item  The Borel transform of the moments from the monomials $w(x)=x^{n}$, with $n>1$, has only one double pole at $u=n+1$, related to the OPE condensate with $D=2(n+1)$ to which the moment is maximally sensitive, in the sense of Eq.~(\ref{OPEdeltaD}).

	\item Moments that are pinched and do not contain the term proportional to $x$ are particularly stable. Their Borel transform is regular at $u=2$ and there is a partial cancellation of the leading UV renormalon, which translates into a smoother series. The series, in these cases, is well described by FOPT while CIPT struggles to approach the Borel sum and runs into the asymptotic behaviour already at $\mathcal{O}(\alpha_s^{4})$ or $\mathcal{O}(\alpha_s^5)$.
\end{itemize}

In the remainder we will discuss the case of QCD. With the use of a convenient scheme transformation and a redefinition of the Borel transform one is able to show that results in QCD are very similar to the ones obtained in large-$\beta_0$.

\subsection{Results in QCD}
\label{resultsQCD}

Two main ingredients enter the discussion of the previous section. First,
we have full knowledge about the renormalon structure of the Adler function. In particular
we know which poles exist and if they are double or simple poles, exactly. Second, in the derivation of Eq.~(\ref{BTdLb2}), because we work in the large-$\beta_0$ limit, we made use of the one-loop running of $\alpha_s$.  In the case of QCD, on the other hand, we have to be content with a partial knowledge about the renormalon structure of the Adler function. The positions of the singularities are unchanged, but now they are no longer poles and become branch cuts. The running of the coupling is also much more involved when terms beyond one loop are included in the $\beta$ function. In order to be able to obtain an analytical expression for the Borel transform of $\delta^{(0)}_{w_i}$, it is useful to leave the $\MSb$ scheme and work in
another class of schemes which have a particularly simple $\beta$ function.

Without loss of generality, we will employ the $C$ scheme introduced in Ref.~\cite{Boito:2016pwf} in the derivation we perform below. The implementation
of the $C$ scheme is based on the fact that, when going from an input scheme, say the $\MSb$, to another scheme that we denote with hatted quantities, the QCD scale parameter $\Lambda$ changes as~\cite{Celmaster:1979km}
\beq
\hat\Lambda = \Lambda_{\MSb}\, e^{c_1/\beta_1},
\eeq
where the coefficient $c_1$ is the first non-trivial coefficient in the perturbative expansion of the coupling $\hat a\equiv \hat \alpha_s/\pi$ in terms of $a\equiv \alpha_s^{\MSb}/\pi$:
\beq
\hat a = a +c_1a^2+c_2a^3+\cdots.
\eeq
With the expression of the scale-invariant QCD $\Lambda$ parameter one can then relate
the two schemes with a continuous parameter $C$, that measures the shift in $\Lambda$, by
\beq
\frac{1}{\hat a_Q} + \frac{\beta_2}{\beta_1} \ln\hat a_Q =  \frac{1}{a_Q} + \frac{\beta_1}{2}\,C +
\frac{\beta_2}{\beta_1}\ln a_Q - \beta_1 \!\int\limits_0^{a_Q}\,
\frac{{\rm d}a}{\tilde\beta(a)},
\eeq
where $C=\frac{-2c_1}{\beta_1}$, we defined
\begin{equation}
\frac{1}{\tilde\beta(a)} \,\equiv\, \frac{1}{\beta(a)} - \frac{1}{\beta_1 a^2}
+ \frac{\beta_2}{\beta_1^2 a},
\end{equation}
 and we have made explicit the renormalisation scale dependence in $a_Q$.
 A relation that is important in the remainder is the analogue of Eq.~(\ref{eq:alphasrunning1loop}) in the $C$ scheme which reads
 \beq
\frac{1}{\ah_Q}=\frac{\beta_1}{2}\ln\left(\frac{Q^2}{\hat \Lambda^2}\right) +\frac{\beta_1\,C}{2}
-\frac{\beta_2}{\beta_1}\ln\ah_Q.\label{eq:oneoverahat}
 \eeq

In this scheme, the $\beta$-function is known exactly and reads
\beq
-\,Q\,\frac{{\rm d}\ah_Q}{{\rm d}Q} \,\equiv\, \hat\beta(\ah_Q) \,=\,
\frac{\beta_1 \ah_Q^2}{\left(1 - \frac{\beta_2}{\beta_1}\, \ah_Q\right)}.\label{betainCscheme}
\eeq
This fact enormously simplifies the task of obtaining a closed form for the Borel transform
of $\delta^{(0)}_{w_i}$ in QCD. Finally, we remark that the dependence on the scheme parameter $C$ is, in fact, governed by the same function 
\beq
-2 \frac{{\rm d}\ah_Q}{{\rm d}C} =\,
\frac{\beta_1 \ah_Q^2}{\left(1 - \frac{\beta_2}{\beta_1}\, \ah_Q\right)}.
\eeq
The coupling becomes smaller for larger values of $C$ and the theory ceases to be perturbative
for $C\approx -1.5$ (using the $\MSb$ scheme as input)~\cite{Boito:2016pwf}.
This means that the coupling in the $C$-scheme depends on a particular combination of the scale and scheme parameters $\alpha_s\equiv \alpha_s(Q^2e^C)$. Scale and scheme variations become, therefore, completely equivalent. The explicit expressions for the perturbative coefficients relating the $\MSb$ and the $C$ schemes, together with further details, can be found in the original publications~\cite{Boito:2016pwf,Boito:2016feb}. Finally, we remark that there is no value of $C$ that corresponds strictly to the $\MSb$, but for
$C\approx 0$ the results are very similar (at one loop, $C=0$ corresponds to the $\MSb$ exactly).

For schemes in which the $\beta$-function takes the form of Eq.~(\ref{betainCscheme}) it is convenient to work with a modified Borel transform defined as~\cite{Brown:1992pk}
\begin{equation}
\mathcal{B}[\widehat D](t) = \sum\limits_{n=1}^\infty \frac{\Gamma(1+\lambdabar t)}{\Gamma(n+1+\lambdabar t)} n \, \frac{\hat c_{n,1}}{\pi^n}
\ t^n, \label{eq:modified-borel-transform-adler}
\end{equation}
where $\lambdabar=\beta_2/(\beta_1\pi)$ and $\hat c_{n,1}$ are the Adler function coefficients in the $C$ scheme.
With this definition, the Borel sum of the series now reads
\begin{equation}
\widehat D(\hat \alpha_s (Q^2)) = \int\limits_0^\infty \frac{dt}{t} e^{-t/\hat \alpha_s(Q^2)}
\frac{[t/\hat \alpha_s (Q^2)]^{\lambdabar t}}{\Gamma(1+\lambdabar t)}
\mathcal{B}[\widehat D](t).
\label{eq:modified-borel-int-adler}
\end{equation}
The asymptotic expansion to the latter result is obtained using Eq.~(\ref{eq:modified-borel-transform-adler}) in (\ref{eq:modified-borel-int-adler}) and gives, as expected~\cite{Brown:1992pk},
\beq
\widehat D = \sum_{n=1}^\infty \hat c_{n,1} \ah^n_Q.
\eeq
The modified Borel transform has renormalon singularites at the same location as the usual Borel transform, but their exponent is shifted. As demonstrated in Ref.~\cite{Brown:1992pk}, if the usual Borel transform has a singularity of the form
\begin{equation}
B[\widehat D](u)\sim \frac{1}{(p-u)^\alpha},
\end{equation}
the modified Borel transforms behaves for $u\sim p$ as
\begin{equation}
\mathcal{B}[\widehat D](u) \sim \frac{1}{(p-u)^{\alpha-\frac{2\pi p}{\beta_1}\lambdabar }},\label{eq:shift}
\end{equation}
with the exponent of the singularity shifted  by $\frac{2\pi p}{\beta_1}\lambdabar =+2p(\beta_2/\beta_1^2)$ and, as before,  $u=\frac{\beta_1 t}{2\pi}$.

Let us now calculate the modified Borel transform of $\delta^{(0)}_{w_i}$ in the C scheme. The calculation is very similar to what was done in large-$\beta_0$. Using Eq.~(\ref{eq:modified-borel-int-adler}) into Eq.~(\ref{eq:delta0}) we obtain
\begin{equation}
\delta^{(0)}_{w_i}=\frac{1}{2 \pi }\int_{0}^{2\pi}d\phi \ W_i(e^{i\phi}) \
\int\limits_0^\infty \frac{dt}{t} \ e^{-t/\hat \alpha_s(-s_0 e^{i\phi})} \frac{[t/\hat \alpha_s(-s_0 e^{i\phi})]^{\lambdabar t}}{\Gamma(1+\lambdabar t)} \label{eq:delta0modborel}
\mathcal{B}[\widehat D](t).
\end{equation}
With the use of Eq.~(\ref{eq:oneoverahat}) one finds
\beq
\text{e}^{-t/\hat \alpha_s(-s_0 e^{i\phi})} = e^{-t/\hat\alpha_s(s_0)}e^{-iu(\phi-\pi)}\left(\frac{\hat \alpha_s(-s_0 e^{i\phi})}{\hat \alpha_s(s_0)}\right)^{\lambdabar t},
\eeq
and inverting the order of the integration in Eq.~(\ref{eq:delta0modborel}) one obtains, for the monomial weight function $w(x)=x^n$, the following result
\begin{equation}
\mathcal{B}[\delta^{(0)}_{x^n}](u) = \frac{2}{1+n-u}\frac{\sin(\pi u)}{\pi u}\mathcal{B}[\widehat D](u). \label{eq:modBoreldelta}
\end{equation}
This shows that the relation of Eq.~(\ref{BTdLb2}) is, in fact, much more general, since any scheme can be brought to the $C$ scheme without loss of generality. The prefactor is the same in QCD and in large-$\beta_0$, and so is the enhancement of the renormalon associated with the contribution with dimension $D=2(n+1)$ in the OPE.\footnote{An approximate relation between the Borel transformed Adler function and the Borel transform of $\delta^{(0)}_{w_\tau}$ can be found in~\cite{Caprini:2019kwp}. The result of Eq.~(\ref{eq:modBoreldelta}) is fully general.} The main difference is that in QCD the singularities of $\mathcal{B}[\widehat D](u)$  are, in general, branch points and are no longer poles.
The exponent of the singularities is related to the anomalous dimension of the associated operator contributing to the OPE.

To make further progress, let us look at the explicit structure of the IR singularities.
In the notation of~\cite{Beneke:2008ad}, the singularities of the usual Borel transform are written as
\begin{equation}
B[\widehat D_p^{\rm IR}] \equiv \frac{d_p^{\rm IR}}{(p-u)^{1+\tilde \gamma }}\left[1+\tilde b_1^{(p)}(p-u)+\tilde b_2^{(p)}(p-u)^2+\cdots    \right],
\end{equation}
where the constants $\tilde \gamma$ and $\tilde b_i^{(p)}$ depend on the anomalous dimension of the associated operator in the OPE as well as on $\beta$-function coefficients.
The explicit expression for the exponent $\tilde \gamma$ is
\begin{equation}
\tilde \gamma = 2p \frac{\beta_2}{\beta_1^2} - \frac{\gamma_{O_d}^{(1)}}{\beta_1}\label{eq:gammatilde},
\end{equation}
where the anomalous dimension associated with the operator $O_d$ is defined as
\begin{equation}
-\mu \frac{d}{d\mu} O_d(\mu) = \left(\gamma_{O_d}^{(1)}a_\mu +\gamma_{O_d}^{(2)}a_\mu^2+\cdots    \right)O_d(\mu).
\end{equation}
For the modified Borel transform we have then
\begin{equation}
\mathcal{B}[\widehat D_p^{\rm IR}] \sim \frac{1}{(p-u)^{1+\tilde \gamma - \frac{2\pi p}{\beta_1}\lambdabar }} = \frac{1}{(p-u)^{1 - \gamma_{O_d}^{(1)}/\beta_1}},
\end{equation}
where the first factor in the r.h.s. Eq.~(\ref{eq:gammatilde}) is exactly cancelled by the shift in the singularity of Eq.~(\ref{eq:shift}). For the discussion of the Borel transformed $\delta^{(0)}_{w_i}$  it is crucial to inspect the leading IR renormalon. This renormalon is related to the gluon condensate which can be expressed in terms of the scale invariant combination $\langle aG^2\rangle$~\cite{Pich:1999hc}.  In this case, the leading IR singularity of the Adler function in the $C$-scheme, and using the modified Borel transform, reduces simply to
\begin{equation}
\mathcal{B}[\widehat D_2^{\rm IR}] \sim \frac{1}{(2-u)},\label{eq:IRCscheme}
\end{equation}
which is a simple pole, exactly as in the large-$\beta_0$ limit. This is remarkable because, with Eq.~(\ref{eq:modBoreldelta}), one can directly translate many of our conclusions from large-$\beta_0$ to QCD, in particular, $\mathcal{B}[\delta^{(0)}_{w_i}]$ has a pole at $u=2$ if and only if the weight function contains a term proportional to $x$. The conclusion that $\mathcal{B}[\delta^{(0)}_{w_i}]$ is less singular also remains valid, and it is again true that the singularity associated with the contributions in the OPE to which the moment is maximally sensitive are not altered by the prefactor of Eq.~(\ref{eq:modBoreldelta}).

In view of the above discussion, and the results of Eqs.~(\ref{eq:modBoreldelta}) and~(\ref{eq:IRCscheme}), we learn that the same mechanisms of suppression, enhancement, and cancellation of renormalon singularities identified in large-$\beta_0$ are also at work in QCD. Similarities between the results in the two cases were identified in Ref.~\cite{Beneke:2012vb} although the explicit connection with the renormalon singularities  of $\mathcal{B}[\delta^{(0)}_{w_i}]$ was not investigated in that work.
Although in QCD we have only partial information about the renormalon singularities, and in particular about their numerator, we are in a position to speculate that the reason
behind the good or bad perturbative behaviour of the different moments is rooted in the
same interplay between the renormalons of $\mathcal{B}[\delta^{(0)}_{w_i}]$.

\begin{figure}[!t]
	\begin{center}
		\subfigure[$\delta^{(0)}$, $w(x)=(1-x)^2(1+2x)$, PAs.] {\includegraphics[width=.49\columnwidth,angle=0]{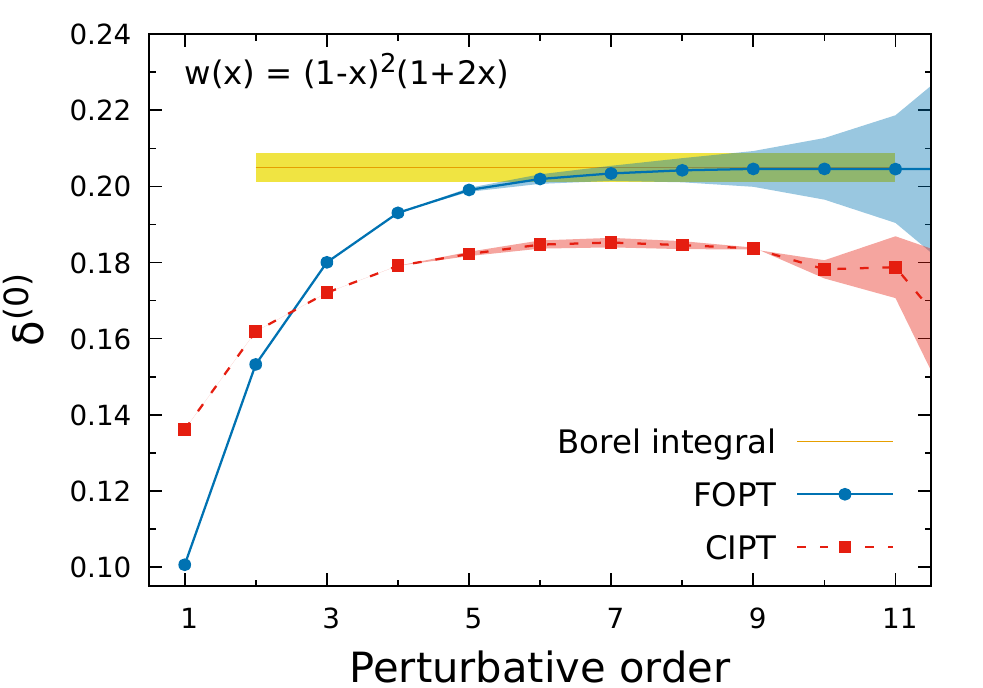}\label{fig:QCDweqtau}}
		\subfigure[$\delta^{(0)}$, $w(x)=1-x^2$, PAs.]{\includegraphics[width=.49\columnwidth,angle=0]{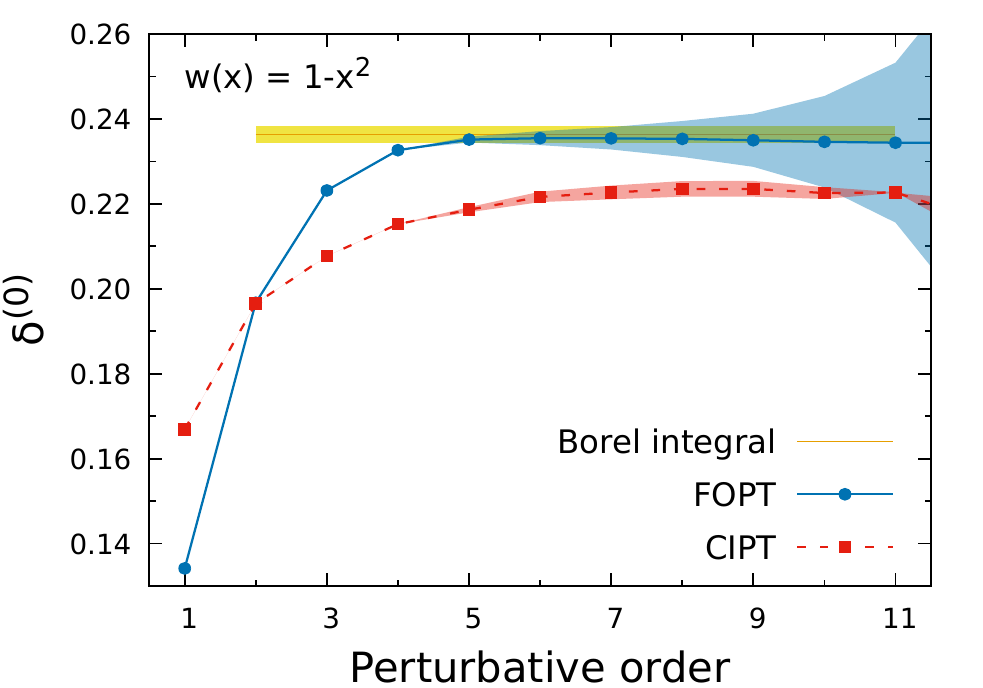}\label{fig:QCDweq1mx2}}
		\subfigure[$\delta^{(0)}$, $w(x)=x$, PAs.]{\includegraphics[width=.49\columnwidth,angle=0]{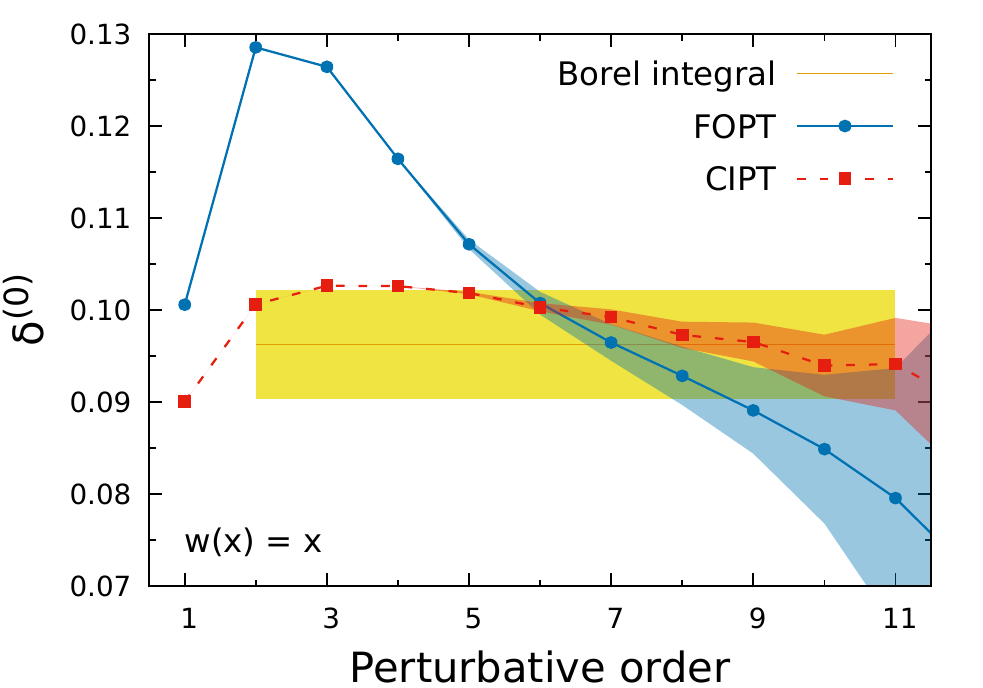}\label{fig:QCDweqx}}
		\subfigure[$\delta^{(0)}$, $w(x)=1-x$, PAs.] {\includegraphics[width=.49\columnwidth,angle=0]{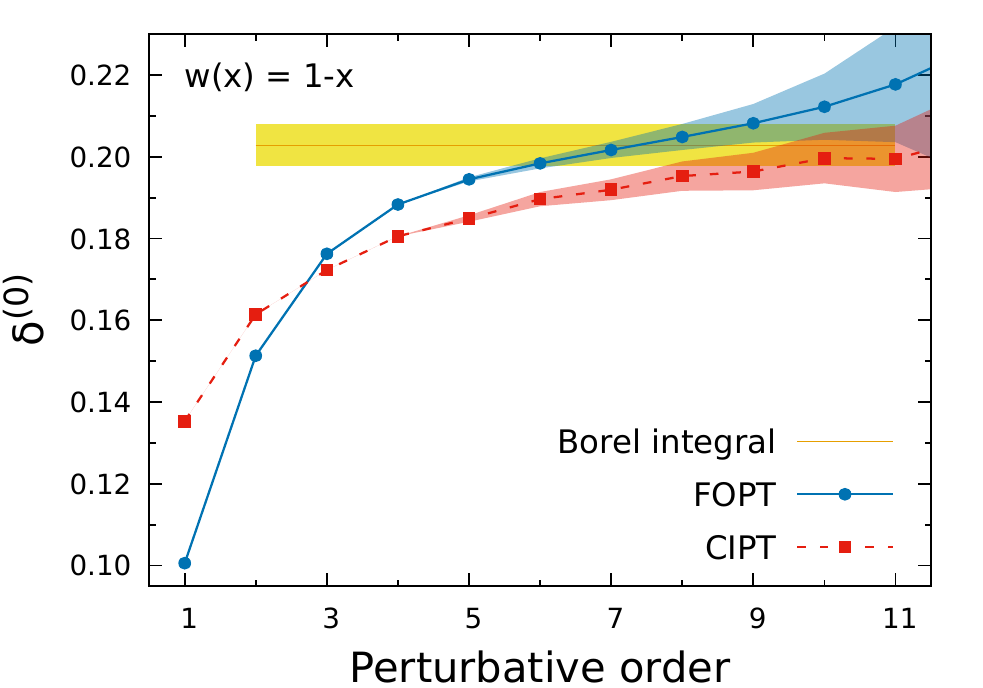}\label{fig:QCDweq1mx}}
		\caption{Perturbative series for four emblematic moments order by order in $\alpha_s$  within the higher-order reconstruction of the QCD perturbative series with the use of Pad\'e approximants of  Ref.~\cite{Boito:2018rwt}, in $\MSb$ and with $s_0=m_\tau^2$.  The shaded bands represent  uncertainties associated with the Pad\'e approximants as discussed in the original reference~\cite{Boito:2018rwt}. In (a) and (b) we show moments with good perturbative behaviour, while in (c) and (d) we display results for moments containing the term  $x$, which show the run-away behaviour of the perturbative series.}
		\label{fig:QCDwPAs}
	\end{center}
\end{figure}

Here, we study the QCD perturbative
series for different moments using the model-independent reconstruction of the higher-order coefficients of Ref.~\cite{Boito:2018rwt}, where the mathematical method of Pad\'e approximants was used to describe the series. In Fig.~\ref{fig:QCDwPAs}, we show results for four emblematic moments, in FOPT and CIPT, with $s_0=m_\tau^2$ and in the $\MSb$; the shaded bands represent  an uncertainty that stems from the Pad\'e approximant method, as discussed in~\cite{Boito:2018rwt}. In Figs.~\ref{fig:QCDweqtau} and~\ref{fig:QCDweq1mx2},
the results for two moments that display good perturbative behaviour are shown: $w_\tau$ and $w(x)=1-x^2$, respectively. The horizontal yellow bands represent an estimate for the Borel integral of the moments within the Pad\'e-approximant description. Again, the FOPT series approaches the true value, as predicted by the Pad\'e approximants, and is rather stable around it until at least the 8-th order. (We relegate to App.~\ref{borelint} a more detailed discussion about the Borel integrals together with a comparison with results from the model of Refs.~\cite{Beneke:2008ad,Beneke:2012vb}, which are similar to ours.)

In Fig.~\ref{fig:QCDweqx}, we show the results for the monomial $w(x)=x$, which exacerbates the
run-away behaviour that stems from the leading IR singularity, as in the large-$\beta_0$  case of Fig.~\ref{fig:TruncatedLbweqx}. Here CIPT is relatively good agreement with the true results, but this is not the case
for other moments containing the linear term, such as $w(x)=1-x$, shown in Fig.~\ref{fig:QCDweq1mx}. This moment inherits the run-away behaviour of the monomial and
both FOPT and CIPT are rather unstable, never stabilizing around the true result.

Finally, one can corroborate our conclusion that the bad perturbative behaviour of moments containing the linear term is related to the leading IR singularity by considering the ``alternative model" of Ref.~\cite{Beneke:2012vb}. In this case, a model for the QCD Adler function is constructed without the leading IR singularity. In the $C$-scheme and using the modified Borel transform, this means that, for this model, the Borel transform of $\delta^{(0)}_{w_i}$
is regular at $u=2$, since in the prefactor of Eq.~(\ref{eq:modBoreldelta}) the pole is cancelled by the zero in $\sin(\pi u)$. As shown in~\cite{Beneke:2012vb}, the run-away behaviour is not present in this case, which shows, once more, that it stems from the leading IR singularity in $B[\delta^{(0)}_{w_i}]$.

In conclusion, with the use of the $C$-scheme and the modified Borel transform, the relation between the Borel transform Adler function and the Borel transform of $\delta^{(0)}_{w_i}$ are formally the same in QCD and in the large-$\beta_0$ limit. The singularities related to the contributions in the OPE are equally enhanced or suppressed and, in general, $\delta^{(0)}_{w_i}$
is significantly less singular than the Adler function. In the case of the leading IR renormalon
the parallel is strict since within this framework it is a simple pole both in QCD and in  large-$\beta_0$. The phenomenological consequences are then the same: moments with a linear term in $x$ display an unstable perturbative behaviour. Finally, the moments with good perturbative behaviour in large-$\beta_0$ are also well behaved in QCD, at least if FOPT is used. The results from the model-independent Pad\'e approximant reconstruction of the series are qualitatively similar to the ``Borel model" of Refs.~\cite{Beneke:2008ad,Beneke:2012vb} which
attests the robustness of our conclusions.

\section{Optimal truncation with scheme variations}
\label{optimization}

We close this work with a  discussion of the optimal truncation of the (asymptotic) series associated with the moments that display good perturbative behaviour. In Ref.~\cite{Boito:2016pwf,Boito:2016yom} it has been suggested that, in the spirit of an asymptotic series, the optimal truncation for the perturbative expansion of the Adler function and of integrated moments is achieved by choosing the scheme (or scale) in which the last known coefficient of the series vanishes.
In this case, by construction, the smallest term of the series, which is zero, is precisely the last known term, which makes it
the ideal point for the optimal truncation of the asymptotic series.\footnote{It is not guaranteed that the truncation at the smallest term, which is known as superasymptotic approximation, is always the optimal truncation, but experience shows that it is very often the case~\cite{Boyd1999}.}  Through this procedure, one expects to make maximum use of the available information from perturbative QCD.
Here we show that this type of optimization works very well in FOPT for all the moments with good perturbative behaviour, within the reconstruction of the series provided by the Pad\'e approximants of Ref.~\cite{Boito:2018rwt}.

We will work in the $C$-scheme and perform variations of the continuous scheme parameter $C$. However, as discussed in Sec.~\ref{resultsQCD}, in this scheme, scale and scheme transformations are essentially equivalent and the same results can be achieved by renormalisation scale variations. 

Let us illustrate the procedure with the help of a concrete case. The FOPT expansion for $w(x)=1-x^2$ and $s_0=m_\tau^2$ in the $C$-scheme is given by
\begin{align}
\delta^{(0)}_{1-x^2} &= 1.333 \ \hat a_Q + (6.186+3C) \ \hat a_Q^2 + (27.77 + 33.17C + 6.750C^2) \ \hat a_Q^3 \nn \\
& + (119.4 + 246.4C + 124.0C^2 + 15.19C^3) \ \hat a_Q^4 \nn \\
& + (90.73 + 1512C + 1329C^2 + 398.9C^3 + 34.17C^4 + 1.333\,c_{5,1}) \ \hat a_Q^5 + \cdots ,
\end{align}
where we show the four exactly known contributions (up to and including $\alpha_s^4$)  plus the first unknown contribution, proportional to $c_{5,1}$. At order $\alpha_s^5$, the terms without $c_{5,1}$ depend only on $\beta$-function coefficients and lower $c_{n,1}$ and are known exactly.  It is customary to include the fifth term in realistic $\alpha_s$ analysis through an estimate of $c_{5,1}$ which here is taken to be $c_{5,1}=277\pm 51$~\cite{Boito:2018rwt} --- but we will see that the results do not depend strongly on the value of $c_{5,1}$. 

\begin{figure}[!t]
	\begin{center}
		\subfigure[$\delta^{(0)}$, $w(x)=1-x^2$.]{\includegraphics[width=.49\columnwidth,angle=0]{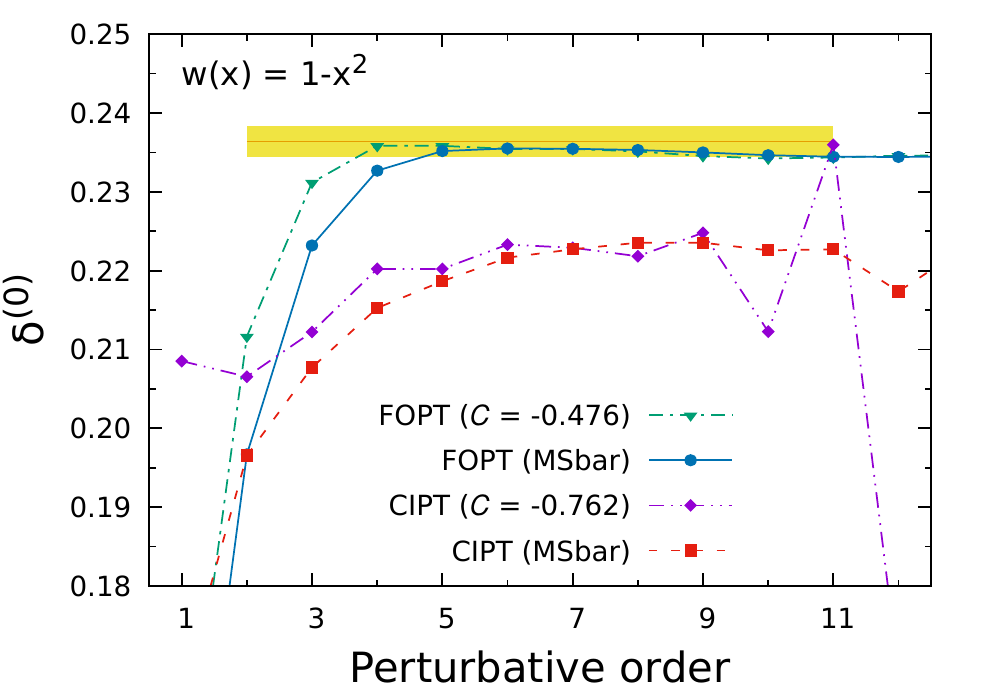}\label{fig:QCDweq1mx2_SV}}
		\subfigure[$\delta^{(0)}$, $w(x)=(1-x)^2(1+2x)$.] {\includegraphics[width=.49\columnwidth,angle=0]{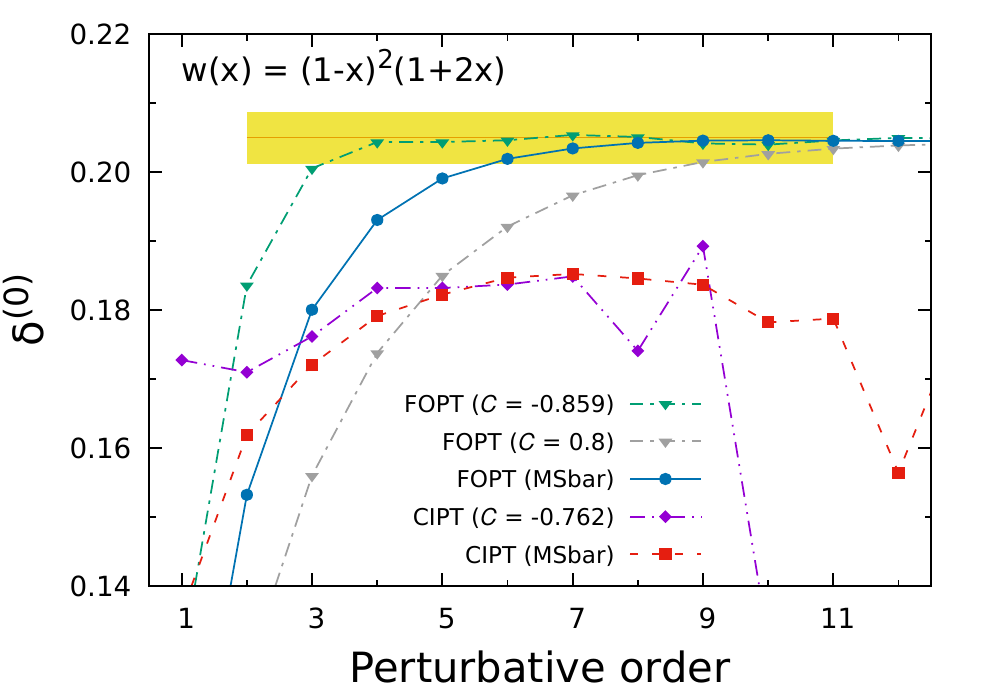}\label{fig:QCDweqtau_SV}}
		\subfigure[$\delta^{(0)}$, $w(x)=1-4x^3+3x^4$.] {\includegraphics[width=.49\columnwidth,angle=0]{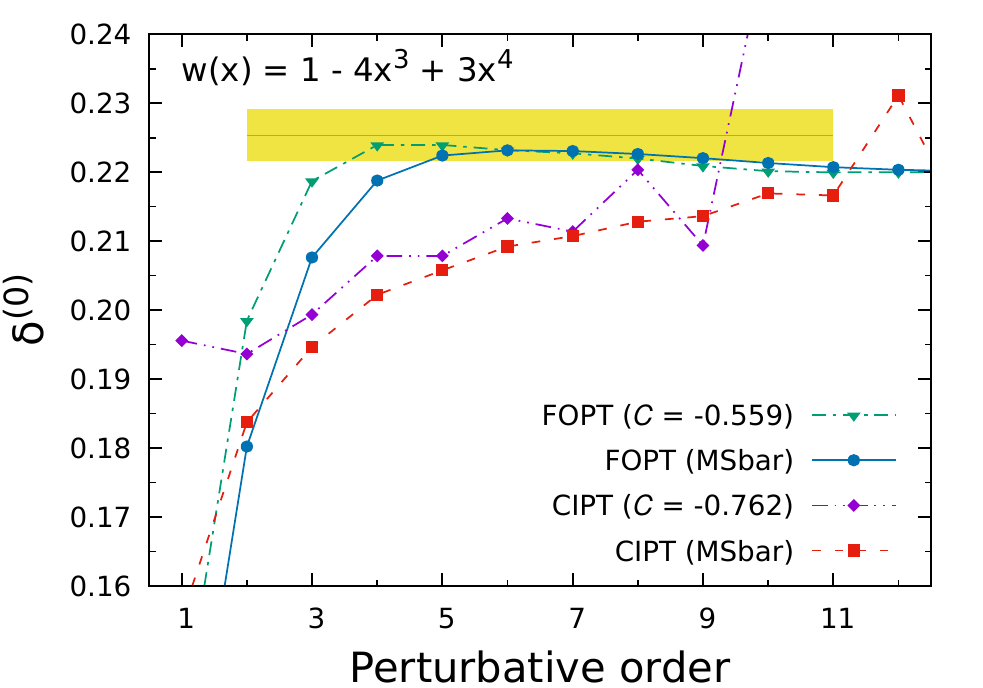}\label{fig:QCDweq1m4x3p3x4_SV}}
		\subfigure[$\delta^{(0)}$, $w(x)=1$.]{\includegraphics[width=.49\columnwidth,angle=0]{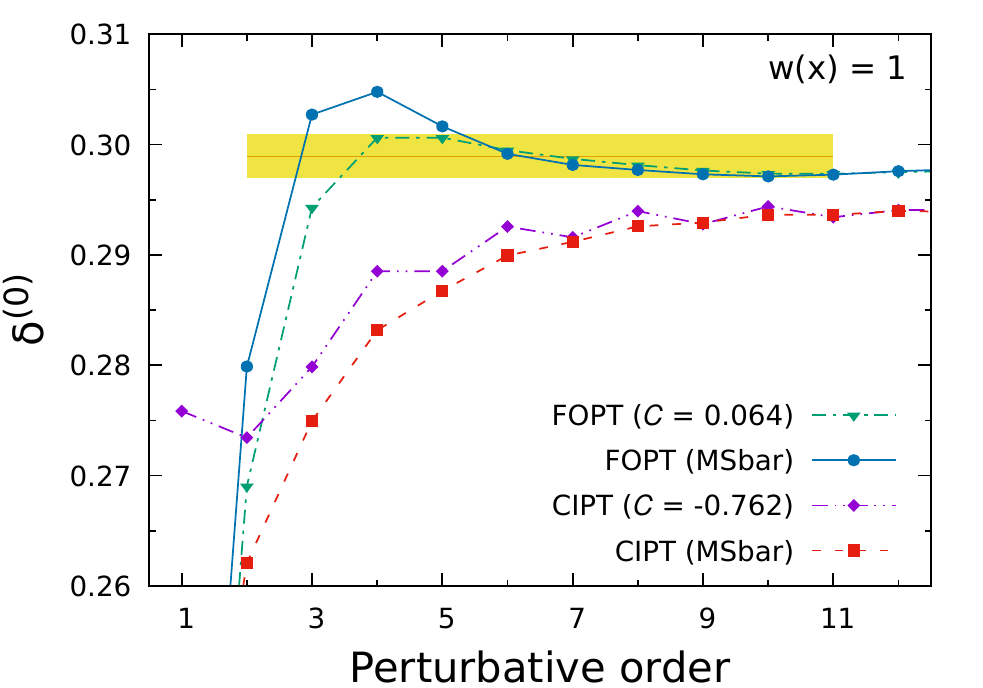}\label{fig:QCDweq1_SV}}
		\caption{Perturbative  series in FOPT and CIPT, for $s_0=m_\tau^2$,  in the $\MSb$ and in the optimal $C$ scheme for three moments with good perturbative behaviour (a), (b), and (c), as well as for the moment of $w(x)=1$, in (d). In (b) the results in grey show a series with large value of $C$, for illustration purposes. The coefficients at $\mathcal{O}(\alpha^5)$ and larger, as well as the Borel sum, are obtained from the description with Pad\'e approximants~\cite{Boito:2018rwt}.  Here we refrain from showing the error bands in each of the perturbative series to avoid cluttering up the plots.  }
		\label{fig:QCDwPAs_SV}
	\end{center}
\end{figure}

The optimized truncation is obtained then by finding the value(s) of $C$ for which the coefficient of $\hat a_Q^5$ vanishes. In the case of $1-x^2$, for FOPT with our central value for $c_{5,1}$, one finds two such values: $C_1= -1.463$ and $C_2=-0.4763$.  The former leads to a rather unstable series, that we discard, since 
the coupling is already entering the non-perturbative regime [$\hat \alpha_s(C_1,m_\tau^2)=0.531$] while the latter is still in the perturbative regime [$\hat \alpha_s(C_2,m_\tau^2)=0.355$] and gives rise to the optimized result. In Fig.~\ref{fig:QCDweq1mx2_SV} we compare the optimized
series for $\delta^{(0)}_{1-x^2}$ (green dot-dashed line) with the usual $\MSb$ result (solid blue line) using the higher-order coefficients and the Borel sum from the description of Ref.~\cite{Boito:2018rwt}. One sees that the optimized FOPT series approaches the true value faster than the $\MSb$ result, already at $\mathcal{O}(\alpha_s^3)$, and remains rather stable around it. This optimization
is related to the larger value of $\hat \alpha_s$ which leads to a series that ``converges" faster than the $\MSb$ one.\footnote{This can be seen as a manifestation of  Carrier's rule: ``Divergent series converge faster than convergent series because they don't have to converge"~\cite{Boyd1999}.} With the optimized series, an estimate of the true result is obtained with the truncation at $\mathcal{O}(\alpha_s^5)$ which gives
\beq
\delta^{(0)}_{1-x^2} (\hat a_{\tau},C=-0.4763) = 0.2358\pm 0.0017,\label{eq:delta1mx2truncated}
\eeq
where the error is due to the variation of $c_{5,1}$ within one sigma.   It is clear from Fig~\ref{fig:QCDweq1mx2_SV} that this leads to an excellent agreement with the true result
--- as predicted from the results of~\cite{Boito:2018rwt} --- which reads $0.2364\pm 0.0020$. 
One should also remark that the procedure is rather independent of the value of $c_{5,1} $ that is used. An uncertainty due to the value of $\alpha_s$, for example, would be about one order of magnitude larger than the uncertainty shown in Eq.~(\ref{eq:delta1mx2truncated}). An attempt to apply 
the same procedure to the CIPT series does not lead to any significant improvement with respect to the (already bad) result obtained in the $\MSb$, as shown in the red  and  purple lines in  Fig.~\ref{fig:QCDwPAs_SV}.

The optimization can also be applied to the kinematic moment, $w_\tau$. The result is again
very good and the acceleration of the series is even more obvious, as displayed in Fig.~\ref{fig:QCDweqtau_SV}. For illustration, we also show, in grey, the series in a scheme with larger value of $C$, namely $C=0.8$ for which $\hat \alpha_s(C=0.8,m_\tau^2)=0.2554$. One sees that in a scheme with a very small value of the coupling the convergence is smooth but very slow for practical purposes, where only
the first few terms are available. Similar results can be obtained for the other moments that have a good perturbative behaviour. As an example, in Fig.~\ref{fig:QCDweq1m4x3p3x4_SV} we show
the result of the optimization of one of the pinched moments introduced in Ref.~\cite{Pich:2016bdg}.

It is also interesting to analyse a borderline case, namely that of  $w(x)=1$. This is not a moment with a bad perturbative behaviour (it does not have a linear term in $x$), but it is also not among the most stable peturbative series, since it does not benefit from the partial cancellation of renormalons. The result in this case is shown in Fig.~\ref{fig:QCDweq1_SV}. Here the $\MSb$ series overshoots the true value   up to $\mathcal{O}(\alpha_s^4)$, as shown in Fig.~\ref{fig:QCDweq1_SV}. In this case,
the value of $C$ that optimizes the truncation turns out positive and the optimal scheme has a smaller value of $\alpha_s$ than in  $\MSb$. The optimization is achieved by avoiding the overshooting of
the true result that is prominent in the $\MSb$ series. The final result is more stable than that in the $\MSb$ and one could expect a smaller error from the truncation of the series, but the acceleration is not very significant. 

Finally, moments with bad perturbative behaviour do not improve in any significant way when we apply the optimization described here. A more stable perturbative expansion  for  these moments can
be achieved with the method of conformal mappings,  making use of the information about
the location of the renormalon singularities~\cite{Caprini:2011ya,Abbas:2013usa,Caprini:2017ikn,Caprini:2018agy}. Even with this technique, in some cases, the series approaches the true value only at high orders.

\section{Conclusions}
\label{conclusions}

In this work, we have discussed in detail the perturbative behaviour of integrated spectral function moments and the connection with the renormalon singularities of their Borel transformed series, denoted $B[\delta^{(0)}_{w_i}]$. The understanding of
the perturbative expansion of such moments is important in guiding the choice of moments employed in realistic
$\alpha_s$ determinations from low-$Q^2$ FESRs. Moments with tamed perturbative expansions are  more reliable and lead to smaller uncertainties from the truncation of perturbation theory.

 In large-$\beta_0$, one can easily establish the relation between
the renormalons of the Adler function and those of the integrated moments in the $\MSb$ scheme. An infinite number of renormalon poles of the Adler function is cancelled 
and $B[\delta^{(0)}_{w_i}]$ is significantly less singular. In particular, for polynomial moments, the leading IR pole is exactly cancelled unless the weight function contains a term
proportional to $x$. The weight functions with this term are therefore the only ones 
that are singular at $u=2$ and they display an unstable perturbative behaviour that
stems from the contribution of this IR pole to the perturbative series.
For the pinched moments that had been identified as having a good perturbative
behaviour in Ref.~\cite{Beneke:2012vb}, we found additional cancellations of renormalon singularities, which are related to a better behaviour at higher orders and postpone the 
asymptotic regime of the series. 

Using the $C$ scheme and a modified Borel transform we have been able to show, in Eq.~(\ref{eq:modBoreldelta}), that the
relation between  Borel transformed moments  and the Borel transformed Adler function is the same in QCD and in large-$\beta_0$. In Eq.~(\ref{eq:IRCscheme}), we have also shown that the leading IR singularity in this framework is again a simple pole. These are the main results of this paper since they
allow us to conclude that the same mechanisms of enhancement, suppression, and partial cancellation of renormalon singularities responsible for the behaviour of the perturbative moments in large-$\beta_0$ are operative in QCD as well. The similar  behaviour of the 
integrated spectral function moments in the two cases is therefore no surprise and again the pinched moments
without the linear term are the best ones (as pointed out in Ref.~\cite{Beneke:2012vb}).
The instabilities related to the leading IR pole are also present in QCD. 

Finally, we have shown that it is possible to use renormalisation scheme (or scale) variations to accelerate the convergence of the moments that display good perturbative behaviour. This had been suggested in Ref.~\cite{Boito:2016pwf} for the $R_\tau$ ratio but
it had never been investigated systematically before. 

In conclusion, we have been able to understand the instabilities and stabilities  of the perturbative expansions of integrated spectral function moments in terms of their renormalons. 
Apart from the implications for the choice of moments in precise $\alpha_s$ analysis, our results can be used in the context of Borel models for the Adler function, since we have shown that scheme transformations and the modified Borel transformed can be used in order to simplify the structure of the leading IR singularity, related to the gluon condensate,
which becomes a simple pole. In fact, the results in large-$\beta_0$ and QCD are therefore much more similar than previously thought. Our findings also suggest that alternative expansions that suppress some of the renormalons may lead to much more stable results, and we plan to 
investigate this issue further in the near future.

\section*{Acknowledgements}

We thank Irinel Caprini, Matthias Jamin, and Santi Peris  for comments on a previous version of the manuscript. Discussions and email exchanges about the modified Borel transform with Santi Peris and Matthias Jamin are gratefully acknowledged.   The work of DB is supported by 
the S\~ao Paulo Research Foundation (FAPESP) grant  No.~2015/20689-9 and by  CNPq grant No.~309847/2018-4. The work of FO is supported by CNPq grant No.~141722/2018-5. This study was financed in part by the Coordena\c{c}\~ao de Aperfei\c{c}oamento de Pessoal de N\'ivel Superior --- Brasil (CAPES) --- Finance Code 001.

\appendix
\section{Conventions for the QCD \boldmath \texorpdfstring{$\beta$}{beta}-function}
\label{betafunction}
We define the QCD $\beta$ function as
\begin{equation}
\beta(a_\mu) \equiv -\mu \frac{d a_\mu}{d\mu} = \beta_1 a_\mu^2 + \beta_2 a_\mu^3 + \beta_3 a_\mu^4 + \beta_4 a_\mu^5 + \beta_5 a_\mu^6 +\cdots,
\end{equation}
where  the first five coefficients are known analytically~\cite{Baikov:2016tgj,Luthe:2017ttg}. It is important to highlight that $\beta_1$ and $\beta_2$ are scheme independent and, in our conventions, they are given by
\begin{equation}
\beta_1 = \frac{11}{2} - \frac{1}{3}N_f, \qquad \beta_2 = \frac{51}{4}-\frac{19}{12}N_f,
\end{equation}
with $N_f$ being the number of flavours. In the particular case of $N_f=3$, relevant here, we have
\begin{equation}
\beta_1 = \frac{9}{2}, \qquad \beta_2=8.
\end{equation}

\section{Details on the Borel integrals from Pad\'e approximants}
\label{borelint}

In this appendix we discuss in further detail how the Borel integrals, or ``true values", of the perturbative series are obtained. We also compare our results with those of Ref.~\cite{Beneke:2012vb}.

Our results are based on the reconstruction of the higher-order coefficients performed in Ref.~\cite{Boito:2018rwt}, using Pad\'e approximants. Several methods have been studied in~\cite{Boito:2018rwt}, using different variants of rational approximants, and constructing the approximants to the Borel transformed Adler function, to the Borel transformed $\delta^{(0)}_{w_\tau}$, as well as to the FOPT expansion of  $\delta^{(0)}_{w_\tau}$. Due to the less singular structure of $B[\delta^{(0)}_{w_\tau}]$, Pad\'e approximants built to this Borel transform converge faster and were the basis for the main result of Ref.~\cite{Boito:2018rwt}, which we use here. The coefficients of the Adler function are then extracted, indirectly, from the series of $\delta^{(0)}_{w_\tau}$.
With these coefficients, given in Tab.~6 of Ref.~\cite{Boito:2018rwt}, one can obtain the expansion of any moment, in FOPT or CIPT, rather accurately up to order $\mathcal{O}(\alpha_s^{10})$.

The main advantage of the use of Pad\'e approximants is that the method is almost completely model independent. In this framework, however, no unique representation of the Borel transformed Adler function is obtained, which makes the task of calculating the Borel integrals for each moment less straightforward. In order to estimate the Borel integrals we have constructed new Pad\'e approximants, following the same methods of Ref.~\cite{Boito:2018rwt}, to each of the Borel transformed $\delta^{(0)}_{w_i}$. In all cases where the moments have good perturbative behaviour, the approximants converge very fast, only three coefficients suffice to obtain a rather stable result. This means that
the prediction from these Pad\'e approximants are based only on the exactly known QCD results. From the Borel transform described by the Pad\'es one can then easily calculate the Borel integral.  Of course, more than one Pad\'e can be built from the same input and we have constructed many different approximants, belonging to different sequences and also using Dlog Pad\'es~\cite{Boito:2018rwt}, in order to estimate the horizontal error band shown in Figs.~\ref{fig:QCDwPAs} and~\ref{fig:QCDwPAs_SV}. The moments containing the linear term $x$, however, lead to less stable results. In order to obtain a stable description of the Borel transform it is necessary to use more coefficients in the construction of the Pad\'e approximants --- which make these results less model independent since they require input from the higher-order coefficients predicted in~\cite{Boito:2018rwt}. The final uncertainties in the horizontal (yellow) bands take into account the dispersion of the results from the use of different Pad\'e approximants as well as the original uncertainty in the prediction of the coefficients of Ref.~\cite{Boito:2018rwt}. In most cases, the former dominates.

\begin{figure}[!t]
	\begin{center}
		\subfigure[$\delta^{(0)}$, $w(x)=(1-x)^2(1+2x)$, PAs.] {\includegraphics[width=.49\columnwidth,angle=0]{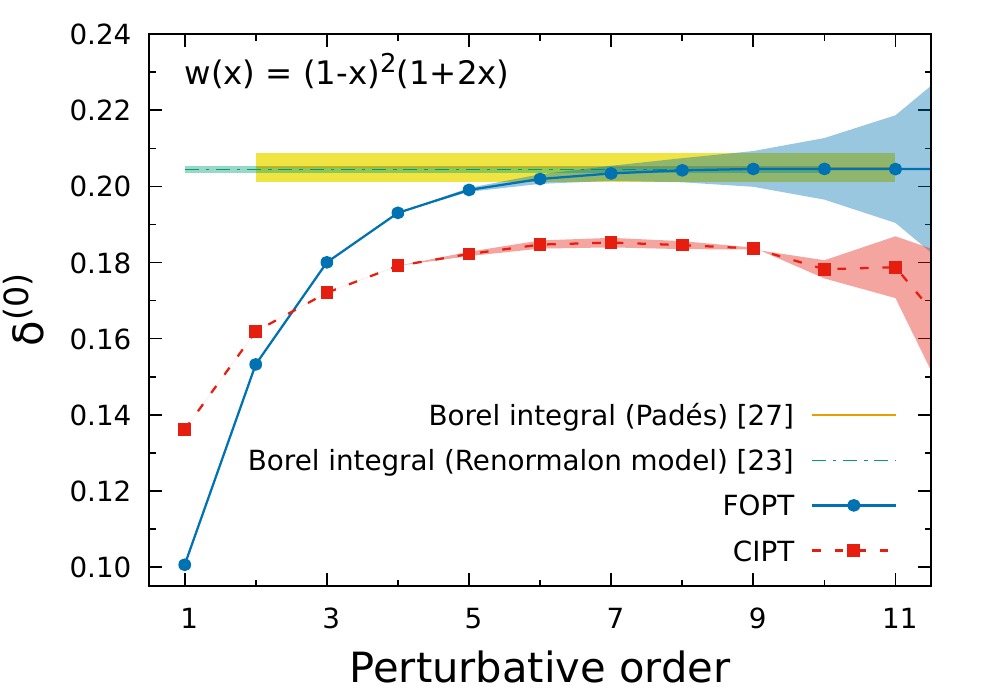}\label{fig:QCDweqtauRM}}
		\subfigure[$\delta^{(0)}$, $w(x)=1-x$, PAs.] {\includegraphics[width=.49\columnwidth,angle=0]{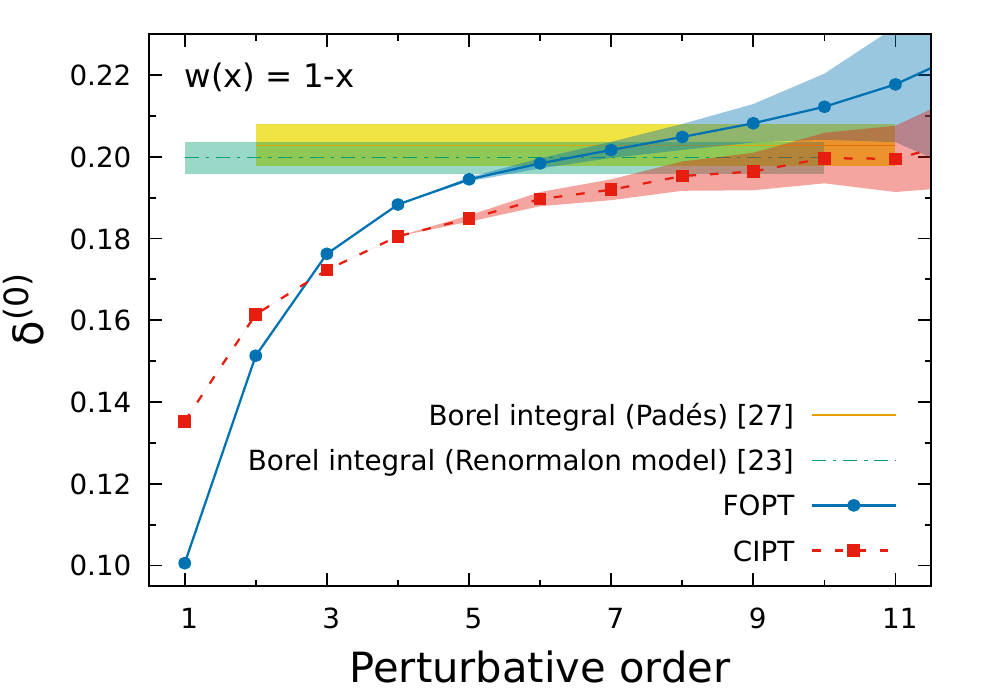}\label{fig:QCDweq1mxRM}}
		\caption{Same as in Figs.~\ref{fig:QCDweqtau} and~\ref{fig:QCDweq1mx}. We add, for comparison, the result for the Borel integral from the ``reference model" of Ref.~\cite{Beneke:2012vb} (green band, with an offset).}
		\label{fig:QCDwPAsRM}
	\end{center}
\end{figure}

Finally, it is interesting to compare our Borel integrals with the ones from the description of Ref.~\cite{Beneke:2012vb,Beneke:2008ad}. In these works,
the Borel transformed Adler function is modelled with its first three dominant renormalons, the leading UV and the first two IR singularities. The residues of the singularities are fixed such as to reproduce the known QCD results. The main advantage of this procedure is that one obtains a unique description of the Borel Adler function, from which all the results are derived. The disadvantage is a possible residual model dependence which could lead to unaccounted systematics. The results from the Pad\'es are, however, in very good agreement with the ``reference model" of~\cite{Beneke:2012vb}, although the uncertainties in the latter case, stemming only from the imaginary ambiguities in the Borel integral, are usually smaller.  In Fig.~\ref{fig:QCDwPAsRM}, we compare the two approaches to the Borel integral, for two exemplary moments, and show that they lead to very similar results. The Borel integral from the reference model of~\cite{Beneke:2008ad} is shown as a green band, with a horizontal offset with respect to the results from Pad\'e approximants, in yellow. In both cases, the FOPT series is preferred. 

\bibliographystyle{jhep}
\bibliography{References}

\end{document}